\documentclass{vgtc}                          

\ifpdf
  \pdfoutput=1\relax                   
  \usepackage{graphicx}                
  \DeclareGraphicsExtensions{.pdf,.png,.jpg,.jpeg} 
\else
  \usepackage{graphicx}                
  \DeclareGraphicsExtensions{.eps}     
\fi%

\graphicspath{{figures/}{pictures/}{images/}{./}} 

\usepackage[colorlinks = true,
            linkcolor = blue,
            urlcolor  = blue,
            citecolor = blue,
            anchorcolor = blue]{hyperref}
\usepackage{microtype}                 
\PassOptionsToPackage{warn}{textcomp}  
\usepackage{textcomp}                  
\usepackage{mathptmx}                  
\usepackage{times}                     
\usepackage{cite}                      
\usepackage{tabu}                      
\usepackage{booktabs}                  

\onlineid{1016}

\vgtccategory{Systems \& Rendering}

\vgtcinsertpkg

\def\code#1{{{\relsize{-1}\texttt{#1}}}\xspace}

\usepackage{enumitem}
\usepackage[capitalise]{cleveref}
\usepackage{adjustbox}
\usepackage{xspace}
\usepackage{relsize}
\usepackage{booktabs}
\usepackage{subcaption}
\usepackage{tikz}

\usepackage{xspace} \newcommand{\FLIP}{\protect\reflectbox{F}LIP\xspace}

\usepackage{color}

\usepackage{stackengine}

\usepackage{xcolor}
\definecolor{dkgreen}{rgb}{0,0.6,0}
\definecolor{dkblue}{rgb}{0,0,0.6}
\definecolor{gray}{rgb}{0.5,0.5,0.5}
\definecolor{mauve}{rgb}{0.58,0,0.82}
\definecolor{commentgreen}{RGB}{2,112,10}
\definecolor{eminence}{RGB}{108,48,130}
\definecolor{weborange}{RGB}{255,165,0}
\definecolor{frenchplum}{RGB}{129,20,83}
\definecolor{darkmagenta}{RGB}{139, 0, 139}
\usepackage{listings}
\newif\ifdiff

\ifdiff
\usepackage{soul}
\def\added#1{{\color{blue}#1}}
\def\removed#1{\textrm{\color{red}\st{#1}}}
\else
\def\added#1{#1}
\def\removed#1{}
\fi

\title{Standardized Data-Parallel Rendering Using ANARI}

\author{
        Ingo Wald\thanks{e-mail: iwald@nvidia.com}\\ %
        \scriptsize NVIDIA %
        \and
        Stefan Zellmann\thanks{e-mail: zellmann@uni-koeln.de}\\ %
        \scriptsize University of Cologne  %
        \and Jefferson Amstutz\thanks{e-mail: jamstutz@nvidia.com}\\ %
        \scriptsize NVIDIA %
        \and Qi Wu\thanks{e-mail: qadwu@ucdavis.edu}\\ %
        \scriptsize University of California, Davis %
        \and Kevin Griffin \thanks{e-mail: kgriffin@nvidia.com}\\ \scriptsize NVIDIA
        \and Milan Jaros\thanks{e-mail: milan.jaros@vsb.cz}\\ %
        \scriptsize IT4Innovations, VSB – Technical University of Ostrava %
        \and Stefan Wesner\thanks{e-mail: wesner@uni-koeln.de}\\ %
        \scriptsize University of Cologne  %
}

\teaser{
  \centering 
  \scriptsize
  
  \resizebox{1.00\textwidth}{!}{
    \setlength\tabcolsep{0pt}
    \stackunder[-8pt]
    {\includegraphics[height=.24\linewidth]{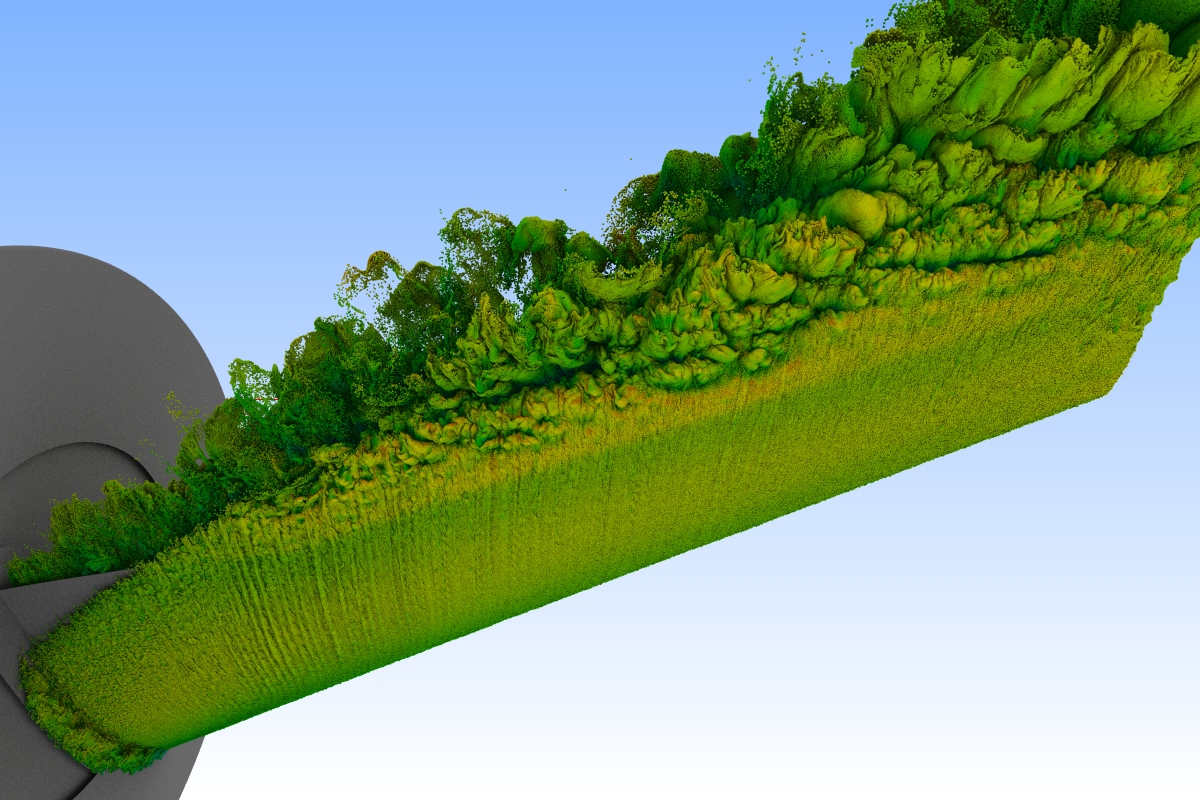}}{(a)}
    \stackunder[-8pt]
    {\includegraphics[height=.24\linewidth]{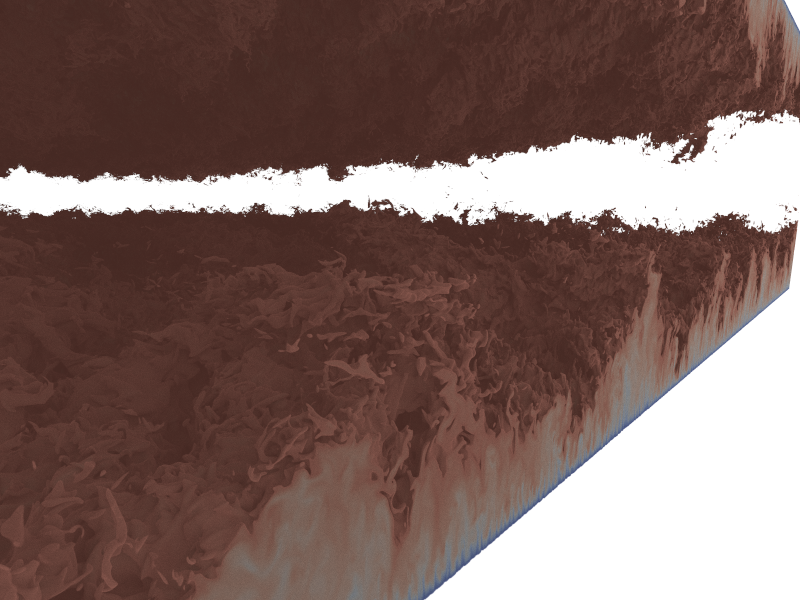}}{(b)}
    \stackunder[-8pt]
    {\includegraphics[height=.24\linewidth]{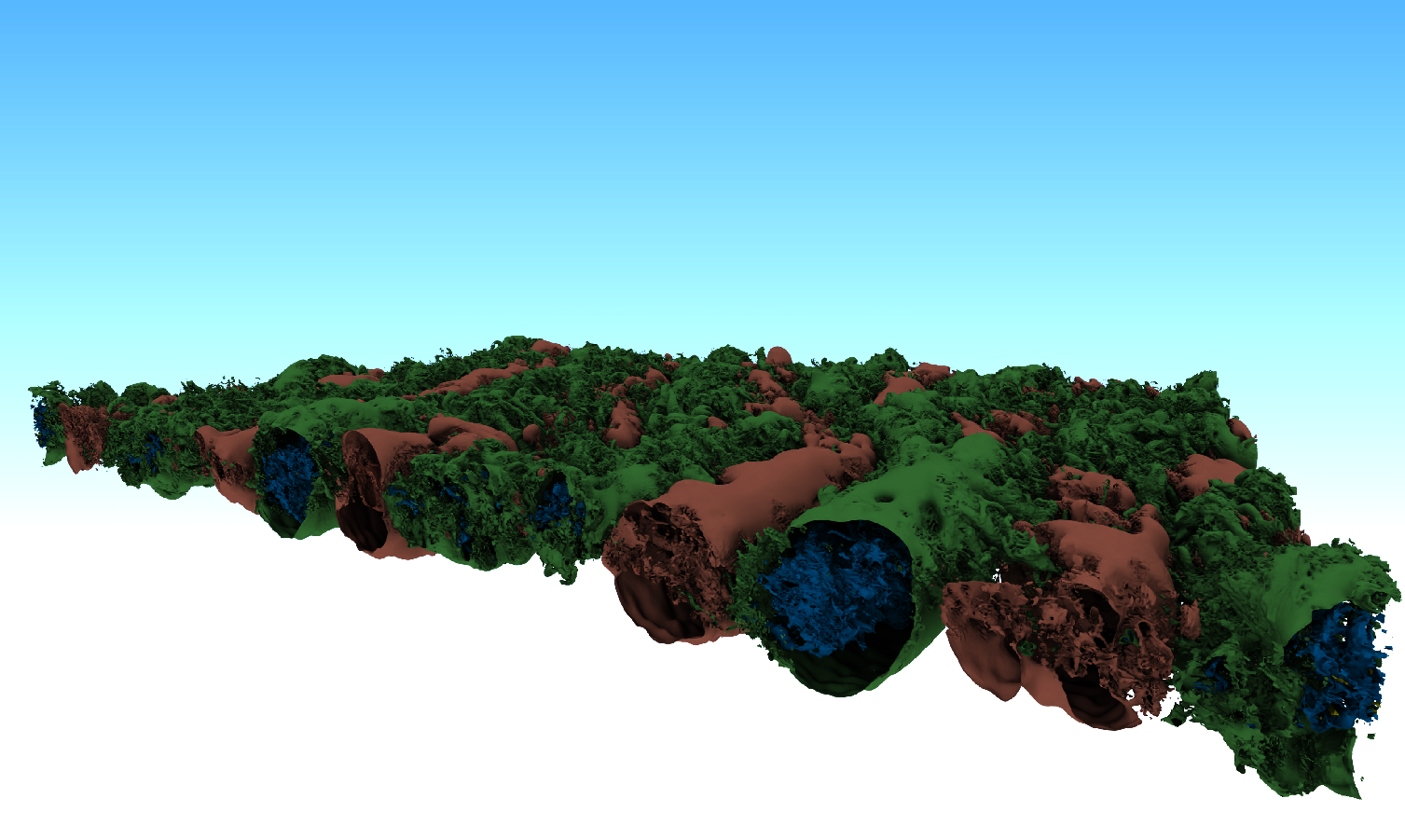}}{(c)}
    \stackunder[-8pt]
    {\includegraphics[height=.24\linewidth]{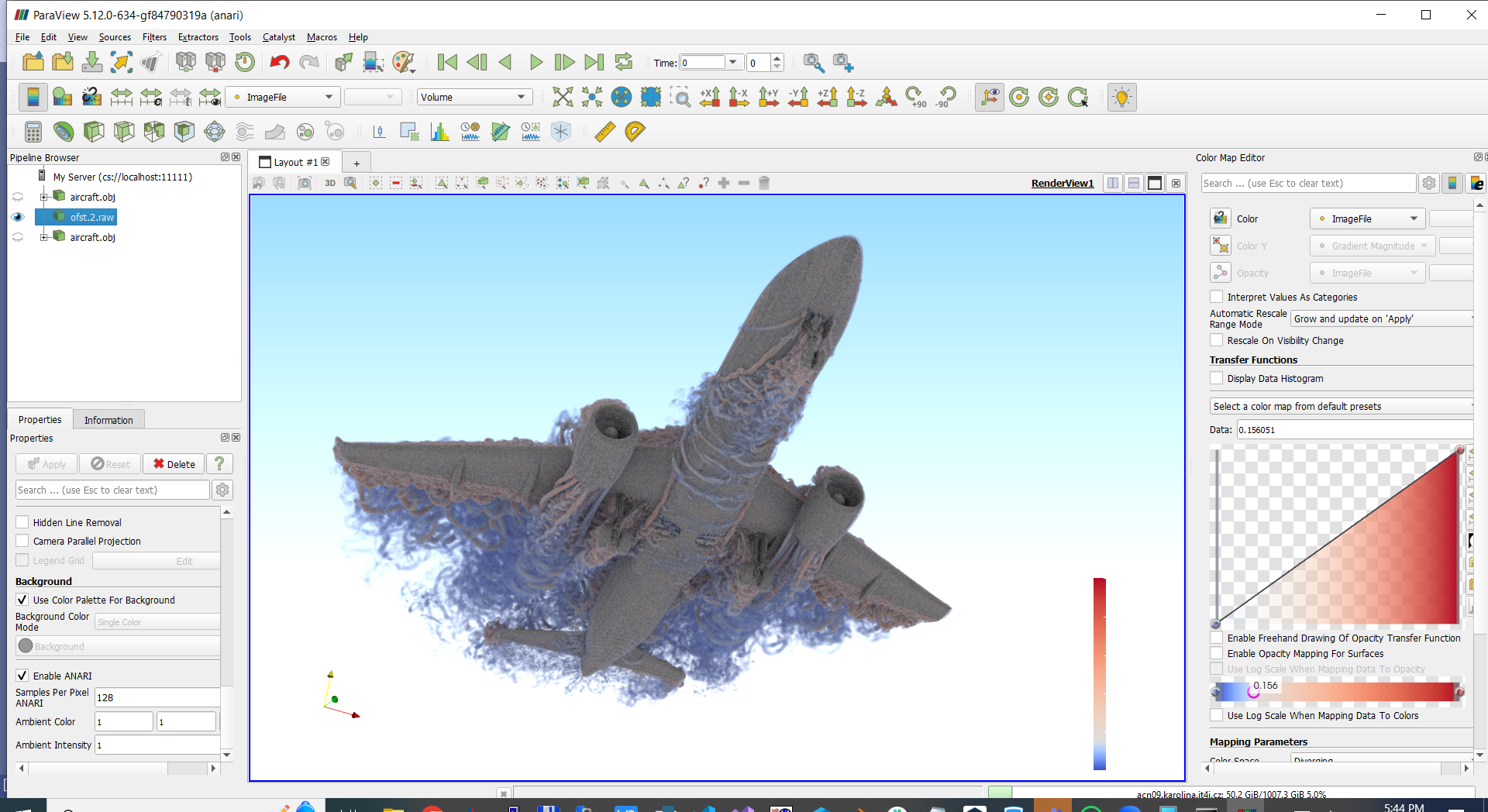}}{(d)}
  }\\[-1em]
  \caption{\label{fig:teaser} Several examples of large sci-vis data
    being rendered using the data-parallel ANARI paradigm proposed in
    this paper. From left to right: a)~Roughly one billion
    color-mapped spheres, rendered using HayStack and BANARI. b)~The
    roughly 500GB \code{DNS} data set, with volume path tracing on 128
    GPUs, also using HayStack and BANARI. c)~An iso-surface rendered
    during an in-situ Ascent session, while attached to an S3D
    simulation. d)~ParaView performing data-parallel rendering
    on the \code{airplane} data set, using our data-parallel ANARI integration in \code{pvserver}.}
  \label{fig:teaser}
  }

\abstract{We propose and discuss a paradigm that allows for expressing
  \emph{data-parallel} rendering with the classically non-parallel
  ANARI API. We propose this as a new standard for data-parallel
  sci-vis rendering, describe two different implementations of this
  paradigm, and use multiple sample integrations into existing apps to show how easy it
  is to adopt this paradigm, and what can be gained from doing so.}
  
\begin{document}


\firstsection{Introduction}
\maketitle

Visualization is about more than rendering, but rendering nevertheless plays a
large role in many vis tools. Rendering is hard: it was already a hard problem
when all such tools could rely on a single common API (e.g. OpenGL); today,
it is further complicated through the emergence of a whole host of different
vendor specific APIs---which for many good reasons vis tools are often loath to
adopt.

To make it easier for vis developers to adopt newer rendering technologies---and
for vendors, to get their tools adopted---the Khronos organization
has  proposed the ANARI API for portable cross-platform 3D rendering~\cite{anari}.
ANARI aims at providing a single API that both vis app developers and
different graphics hardware vendors can all agree on, providing
important benefits to both sides. For vis app developers, it 
\removed{means that they can target a single API, without having to adopt
vendor-specific APIs; for those developing the ``back-ends'' that implement
this API it means that technologies have been put into that implementation.}
\added{means they can target a single API without having to adopt vendor-specific
APIs, and without compromising the availability of state-of-the-art rendering
features. For those developing ANARI implementations, it means higher adoption
rates and faster adaptation in technological changes as the back-ends become
directly available to the vis apps.}

\added{To the app developer, ANARI presents itself in a pragmatic way: ANARI is an API
allowing the user to specify scene data and rendering frames. Content changes
are expressed by updating objects involved in rendering, such as cameras or data
arrays containing geometry, materials, colors, etc. These objects ultimately
represent a generic interface to the private implementation of the back-end,
where the mechanics of rendering frames is left up to the implementation.}

ANARI is not a silver bullet, though. Even with a single agreed-upon API,
different implementations can and will still differ in what features exactly
they will support (and in which form). Thus, applications still need to be
aware of which specific implementation they may be running on---and either adopt
a least common denominator approach, or have some application features only
available from specific ANARI vendors. Still, this standardization is
encouraging as ANARI is already seeing adoption even in VTK and VTK-m, and
through that, in a variety of tools that use
these~\cite{paraview,visit,vmd,ovito}.

It is paramount to observe that ANARI currently does \emph{not} explicitly
cover data-parallel rendering. This does not mean that data-parallel
applications such as ParaView/Catalyst or VisIt/libsim cannot use ANARI for
their own data-parallel rendering. They certainly can by using ANARI to render
on each node, and then compositing the resulting per-rank images in exactly the
same way they have always done, using image-compositing libraries such as IceT.
This approach is already used in practice, but is intrinsically limited to
whatever image-compositing can or cannot achieve.

\removed{How exactly such a
renderer would then render its image (e.g., by fetching data from
remote nodes, or by sending rays to those nodes) would still be up to
compute global effects in the correct way. This, in fact, is a
puzzling conundrum at the heart of sci-vis today: sci-vis is already
adopting ANARI, with the goal of standardizing around a single
rendering API for ray tracing---yet current ANARI cannot properly
express the kind of data-parallel rendering that high-end sci-vis
requires.}

Rendering intermediate images independently on each node, sorting them in order
and compositing them \emph{at the end} only results in correct images for the
most simple shading models. If we are interested in any global effects---even
such rudimentary effects as shadows or ambient occlusion---an individual surface
cannot be shaded without knowledge of neighboring surfaces; and such neighboring
surfaces can, and in general do reside on other compute nodes. This fundamental
problem can not be solved merely by using ghost and halo regions, as lighting
effects are a global phenomena.

\added{Such distributed renderers relying on ray tracing solve the problem efficiently
by queuing and exchanging rays not at the end of rendering a frame, but instead
constantly while the image is generated. GPU renderers based on ray wave-fronts
are demonstrably efficient at doing this making the approach considered
state-of-the-art, allowing sci-vis tools to benefit from higher-quality
rendering techniques at very little extra cost compared to standard
data-parallel sci-vis renderers.}

\added{What we also observe today is that sci-vis tools like Paraview or
VisIt \emph{do} adopt those higher-quality rendering effects, e.g., by
integrating OSPRay or VisRTX in their standard rendering pipelines; but
when visualizing larger data sets requiring distributed rendering, even when
using those APIs, have to resort to local shading only because when combined
with ordinary image compositing cannot create artifact-free images otherwise.
This, in fact, is a puzzling conundrum at the heart of sci-vis today: sci-vis
is already adopting ANARI, with the goal of standardizing around a single
rendering API for ray tracing---yet current ANARI cannot properly express the
kind of data-parallel rendering that high-end sci-vis requires.}

In this paper, we explore and argue for the concept of \emph{data-parallel
ANARI}. We do so through the lens of using the existing ANARI API to
define---for certain types of data parallel ANARI devices---a \emph{distributed
world} whose individual consistent parts are located on different
\emph{collaborating} ranks. We show that this can be done by merely defining the
\emph{semantics} of a data-parallel ANARI device, and how the different ranks'
individual API calls will \emph{jointly} define such a common distributed world.
We then describe two different sample implementations that each implement this
paradigm---with very different capabilities and limitations---and show the
potential of this approach using a set of diverse applications that make use of
these implementations. In particular, we show several examples of what this
explicitly distributed data-aware paradigm can realize that a purely
application-side compositing-based approach can not.

\section{Background and Related Work}

\subsection{Data Parallel Rendering}

Sci-vis applications can be classified as either post-hoc or
in-situ/in-transfer, where in the latter case the visualization and
analysis pipeline is executed as the data is generated. Post-hoc is
the alternative processing paradigm to in-situ/in-transfer, where data
is saved to permanent storage and then loaded by the sci-vis
application to perform visualization and analysis.

At scale, the dominating operation of a traditional sci-vis renderer
that uses rasterization, simple shading, etc., is sort-last image
compositing; the literature has focused on optimizing this operation
through efficient algorithms such as parallel direct send~\cite{pds,pds2}
or radix-\textit{k}~\cite{radixk}. These developments culminated in
the IceT image compositing library~\cite{icet} which has become the
de-facto standard for real-time distributed renderers. The distributed
rendering back-ends of visualization packages like VisIt~\cite{visit}
or ParaView~\cite{paraview}, as well as in-situ visualization frameworks
like libsim~\cite{libsim}, Catalyst2~\cite{catalyst2}, or
Ascent~\cite{alpine,ascent} all internally build on IceT.

The implication of using sort-last is that compositing happens at the
end, once each rank has reduced its rendering operation to a single
fragment (color, opacity, and depth) per pixel. In reality, that
restricts the application in what kind of content it can render in
this way; in particular, compositing with a single fragment only works
correctly if data is partitioned in a way where each rank's data is
convex.  To remove some of these restrictions researchers have looked
at various forms of ``deep'' frame buffers that can store more than
one fragment per pixel, which in turn allows for better handling
transparency when data is partitioned in non-convex ways. A very early
example of this was proposed by Ma~\cite{Ma:unstructured}, more recent
ones by Binyahib et al.~\cite{Binyahib:TVCG} and Sahistan et al.~\cite{big-lander}. The same concept is also used by
OSPRay~\cite{ospray}, whose data-parallel rendering mode relies on
using a \emph{distributed frame buffer}~\cite{usher_scalable_2019}
where ranks can also produce more than one fragment per pixel.
\added{One an interesting contradiction becomes obvious when looking
at OSPRay: while in non-data parallel OSPRay supports advanced rendering
techniques realized with Monte Carlo ray tracing, when rendering in
parallel within ParaView or VisIt, OSPRay has to switch to simple ray
casting without secondary rays, resulting in similar quality as the
typical rasterization renderers used by other sci-vis applications.}

\subsection{Data Pararallel Ray Tracing}

For ray tracing, data parallel rendering means that any ray on any
rank may at any time need intersection with geometry stored by any
other rank. This can be done by using either one of two alternative
techniques: fetching the remote data to the rank that traces the ray
(typically involving some caching of said data,
e.g.~\cite{ize-caching,demarle-caching}), or sending the ray to the
rank that has the data (see, e.g.,~\cite{hypercube-rt,spray,galaxy,stanford-r2e2,brix,rqs}, or
some combination thereof (e.g.,~\cite{reinhard::PhD,navratil,navratil::PhD}).
%
%


Though some of these approaches have been around for decades, this
kind of data parallel rendering has only recently seen interest from
the sci-vis community, probably because only recently hardware has
become capable enough to actually do this. In particular,
SpRay~\cite{spray}, Galaxy~\cite{galaxy}, BriX~\cite{brix}, and
RQS~\cite{rqs} are all been shown to achieve interactive performance
for non-trivial data.

\added{Data parallel CPU and GPU ray tracers aimed at sci-vis achieve
illumination effects, including shadows or ambient occlusion, but also global
illumination with diffuse reflection, by queuing and exchanging rays across
ranks~\cite{galaxy,brix,rqs}. Shading effects are computed by light rays
``bouncing'', i.e., they change direction upon interaction with surfaces,
volumes, or lights. Each time a ray bounces it consequently has to visit scene
content in completely different spots, resulting in highly incoherent patterns.}

\added{To avoid latency, when exchanging rays across ranks, data parallel ray
tracers do that in batches rather than exchanging small groups or
even individual rays. A core concept for that is the ray wave-front: instead of
computing the recursive ray tracing function per each ray individually, compute
kernels are executed per bounce, on all the rays the rank is currently
responsible for. After a bounce finished, the ray wave-fronts are synchronized,
can optionally be sorted and compacted; this is also the point where a
\emph{data parallel} wave-front ray tracer will synchronize its wave-fronts
across ranks using MPI broad-casts or uni-casts. The ranks exchange the rays
they are responsible for until a maximum number of bounces was reached, until
all wave-fronts contain zero rays; etc. The main difference of state-of-the-art
methods these days is how the renderers compute, cull, and assign ray
wave-fronts to ranks in-between bounces.}

\added{Traditionally, the reasoning when distributing rays was to reduce the
overall bandwidth. Optimizing ray tracers account for that by building
complex culling accleration structures~\cite{galaxy,brix} to minimize the overall
number of rays exchanged. Those data structures can be extremely unbalanced,
and especially if the data itself is instanced requires to add additional
constraints to avoid overflowing memory~\cite{zellmann-egpgv-2020}.}

\added{An alternative approach, also adopted in this work, by the data
parallel ANARI back-end described in \cref{sec:barney}, is to distribute the
scene data in a very simple way---e.g., round-robin, bin-packing until all
GPU memory is used, etc.---and then just generate full wave-fronts on all
ranks. The ranks can initially discard some of the rays in their wave-fronts by
testing them for visibility against their geometry, but still generally exchange
more rays than with an optimized partitioning; wave-fronts are exchanged
in a ring buffer or similar pattern. The reason that this works and is not
overly costly is that wave-fronts are sent in lock-step, i.e., all the
communication happens at the same time and is then limited by the pair of ranks
with the highest ray count to exchange. When adding more ranks, the overall ray
count stays the same, resulting in more ranks exchanging fewer rays
individually. This concept was evaluated by Wald et al.~\cite{rqs} on multi-GPU
systems with NV-Link interconnect, but also extends to massively parallel
architectures like GPU clusters. One compelling reason for preferring this
method over approaches that optimize for bandwidth is its negligible
pre-processing time; whereas building an optimized acceleration structure can
often take hours and is then impractical for visualization in a prodcution
context.}

\subsection{ANARI}

ANARI is a cross-vendor 3D rendering API maintained by the
Khronos Group. ANARI connects applications from diverse domains 
to any 3D rendering engine implementing the ANARI API
while still giving implementations a vast degree of freedom of
how exactly rendering is done.
%
%
ANARI has already been adopted by major scientific visualization
 packages, namely VTK~\cite{vtkBook}, VTK-m~\cite{vtkm},
ParaView~\cite{paraview}, VisIt~\cite{visit}, VMD~\cite{vmd}, and OVITO~\cite{ovito}.
ANARI is not limited to scientific visualization, and has also
been integrated, for example, in Blender~\cite{blender}, and OpenUSD's
Hydra subsystem~\cite{hydra-talk}.

\added{A comprehensive introduction to ANARI is out of the scope
of this work; instead we briefly summarize what we consider the main components
of the API that are relevant for data parallel rendering. For a complete
overview we refer the reader to~\cite{anari}}.

ANARI's core design centers around opaque handles to objects representing
the various bespoke actors commonly found in rendering an image---surfaces,
materials, volumes, cameras, lights, frames etc. These objects are
parameterized through generic parameters represented by name/value
pairs and are transitioned between states using parameter commit
semantics---specifically that committing the object's parameters
indicates that those changes should be visible in the next rendered frame.
Alas, frames are rendered asynchronously, where the application
triggers a render operation to start, and then is free to
synchronize with it to access resulting output buffers.

The foundational object for most API calls is the \emph{device}, which
represents the instance of the rendering engine handling ANARI API calls.
After creating a device, applications make all ANARI API calls through
this special handle, which provides implementations a point to
reconcile any common implementation-wide state and gives applications
a clear set of rules for how implementations can be used concurrently.
Thus for the rest of this paper, the phrases ``ANARI implementation''
and ``ANARI device'' will be used synonymously.

\added{Once the ANARI device was initialized, a valid \code{ANARIFrame}
object is all that is required for rendering. A frame object has
color, depth, and other optional auxiliary memory buffers allocated that
it can retire rendered pixels to. A valid frame has a \code{ANARIRenderer}
object, a \code{ANARICamera}, and \code{ANARIWorld} assigned to it.
The ANARI world defines an immediate mode scene graph. It is a special
group node serving as a collection of surfaces, volumes, and light objects.
In addition to that it may also contain \code{ANARIInstance} objects that
can themselves contain group nodes; \code{ANARIWorld} is a special group node
that can contain instance nodes; i.e., the the scene graph has depth two at
most (two-level hierarchy). Buffer objects for input and output are realized as
\code{ANARIArray}s. Arrrays deviate from the otherwise strict immediate mode
model in that they can be mapped and their content altered by the user.}

\section{Data Parallel ANARI (DP-ANARI)}

\removed{The main observation driving this paper is that there is currently one
large contradiction at the heart of sci-vis rendering: On one hand, the
sci-vis community seems to have adopted the goal of going towards ray
tracing, and seems to have chosen ANARI to reach this. On the other,
at the high end sci-vis rendering \emph{needs} data-parallel
rendering, but the ANARI API today is purely single rank, and the
community's existing approach to data parallelism---sort-last
compositing of individual ranks' frame buffers---for ray tracing will
not work. The goal of this paper is to fix this.}

\added{The goal of this paper is to propose a paradigm for data parallel
rendering using ANARI that fits the object model of a data parallel ray tracer
without the limitations of image compositing.} One option would be to
propose a completely new API, but this would be asking the vis
community to completely discard all existing ANARI progress, and start
anew. Instead, we decided to look into what it takes to extend the
\emph{existing} ANARI API to also do data parallel rendering. As it
turns out, this can actually be done without \emph{any} additional new
API calls, by simply proposing a new set of semantics of what the
calling of different API calls on different ranks actually means.

The core of our work is defining the semantics of ANARI API usage in the context of
a distributed application environment. Specifically, all ANARI API calls are usable
as-is, but have additional semantics and constraints applied to them. The following
subsections will outline these semantics and constraints.

\subsection{Object Locality and Consistency}

A concept that occurs in distributed rendering is how objects relate to one
another between nodes. For some objects, they are only defined and interacted
with on the node in which they are created, while for others there must be a
global definition to them on every rank. The following describes the different
application of object definitions with respect to their global or local definitions.

\subsubsection{Globally Consistent Objects}

Some objects are considered \code{global} in the sense that they represent a single,
cooperative entity in the ANARI object hierarchy on all ranks. We define all
\code{ANARIFrame} and \code{ANARIWorld} objects to be considered global objects
where their global identity is established by the order of their construction. Thus
all ranks which use an \code{ANARIFrame} an \code{ANARIWorld} handle must have those
objects constructed as their respective N'th object of that type, effectively requiring
all ranks to create the same number of these objects.

Some objects must have their parameters match on every node in order to have
a well-defined image---\code{ANARIFrame}, \code{ANARIRenderer}, and
\code{ANARICamera}. Applications must use the same sub-type and parameterize
these objects identically, otherwise the output of the resulting image will
be undefined.

\subsubsection{Locally Defined Objects}

All objects under \code{ANARIWorld} are locally defined within the rank on which
they are created and are globally visible (i.e. secondary illumination when
applicable) in the final rendered image. This includes anything which can be
contained with the \code{ANARIWorld} -- instances, groups, surfaces, volumes,
geometries, materials, spatial fields, samplers, and even arrays themselves.
There is no application requirement that any object within the world has any
knowledge or connection to an object on any other rank.

\subsubsection{Locally Mapped Frame Buffers}

One seemingly innocuous question is how and where the data parallel
application can actually access the pixels that a distributed
\code{anariRenderFrame} call has produced. This sounds like a trivial
problem, but is not: it is easy to imagine some apps wanting to map the
entire frame buffers on all ranks, or for others to have different
ranks map different regions of a frame, or ranks just having some
call-back mechanism for image tiles that a given rank has produced
(see, e.g.,~\cite{usher_scalable_2019,han_displaywall_2020}). On the other
hand, trying to capture all these potential use cases would not only
require significant extensions to the API, but also raise the
cost for device developers to implement all these options.

For this trade-off, we intentionally opt for simplicity over flexibility, and
specify that data parallel ANARI devices are only responsible for providing the
final frame buffer on rank~0\footnote{This refers to rank~0 of the MPI
communicator used to initialize the data-parallel ANARI device; if the app
wants this to be on another rank than \emph{its} rank~0 it can of course use
MPI's \code{split} operation to create a new communicator for this.}: Though
all ranks participate in specifying and rendering the scene only a single
rank---rank~0---contains the final image buffers (designated by a parameter on
the participating \code{ANARIFrame}); and on this rank the application can map
this using the existing \code{anariMapFrame} in exactly the same way a
non-parallel application would have. 

While it is not an error to map the frame on another rank, its
dimensions, pixel type, and buffer contents are undefined. Given that
this operation is inherently local, mapping frame buffer outputs has
no synchronization requirements across nodes. ANARI's
asynchronous frame rendering semantics however still apply as they do in
traditional single-node rendering setups. An added benefit of choosing
this route is that there is no difference whatsoever in how
a data-parallel app maps a frame buffer on a data parallel device, vs.\
how a traditional single-process application does on a non-parallel
device (also see \cref{sec:discussion-single-rank}). Of course,
any use cases not captured through this paradigm could still get added
later on through ANARI's extension mechanism.

\subsection{Collaborative Operations}

Most ANARI API calls can be done independently on each
rank, but a few will behave as a rank synchronizing operation---in other words,
some ANARI API calls will necessarily require all ranks to participate in and
will implicitly barrier at that call.

The first and most obvious synchronizing API call is
\code{anariRenderFrame}, as it is the central place that all object parameter
transactions must be completed and where the vast majority of the
implementation's work is done. While the mechanics of rendering a frame is
intentionally left as implementation defined, applications must follow this
semantic to express that every node is ready for its local \code{ANARIWorld}
is ready to render.

Similarly, \code{anariGetProperty} is a synchronizing API for objects that
have a global identity (\code{ANARIFrame} and \code{ANARIWorld}). This is
to ensure that implementations can guarantee global consistency to those
objects when properties are queried by the application. However, this
constraint is only required when applications pass \code{ANARI\_WAIT} to
\code{anariGetProperty}, as \code{ANARI\_NO\_WAIT} would indicate it is
permissible for the device to return a previously held value (or none at all)
in order to prevent blocking.

Finally, \code{anariRelease} is a synchronizing API when called on the
\code{ANARIDevice} which has no remaining application references. This permits
implementations to rely on the device being destroyed in lock-step and
guarantee no additional ANARI API calls will be made using the device.

\subsection{How this works out in practice}
\label{sec:illustration}

\removed{We observe that though this paper is about a more formalized way of
\emph{thinking} about such a data-parallel paradigm---and more clearly
defining how exactly it is supposed to work---in practice it actually
matches very well how existing data parallel applications work. For
example, consider an application that already does parallel rendering,
for example via IceT image compositing: IceT also requires collective
MPI communication, so an application would \emph{already} have a
synchronization point where all ranks need to enter rendering.
In between these ``collective'' render calls the app would be free to do with
its local data as it pleases, and while rendering, different ranks would likely
operate independently, with different ranks rendering different content using
different numbers of calls to whatever library it uses for per-rank rendering.}
\added{We observe that this new data-parallel paradigm in practice actually
matches how existing data parallel applications work. An application using IceT
for compositing would also use MPI for synchronization. The individual ranks
would operate largely independent of each other and render different content
using calls to whatever library the app uses for per-rank rendering.}
Similarly, that application would also have to ensure consistency between
different ranks' global rendering information like camera, background color,
etc., so would likely already have some mechanism to synchronize such
information before it calls rendering. For our paradigm, this is exactly the
same, except that we apply it to ANARI, and formalize the process.

\begin{figure}[ht]
\begin{center}
 \resizebox{0.7\columnwidth}{!}{
  \includegraphics[width=.96\columnwidth]{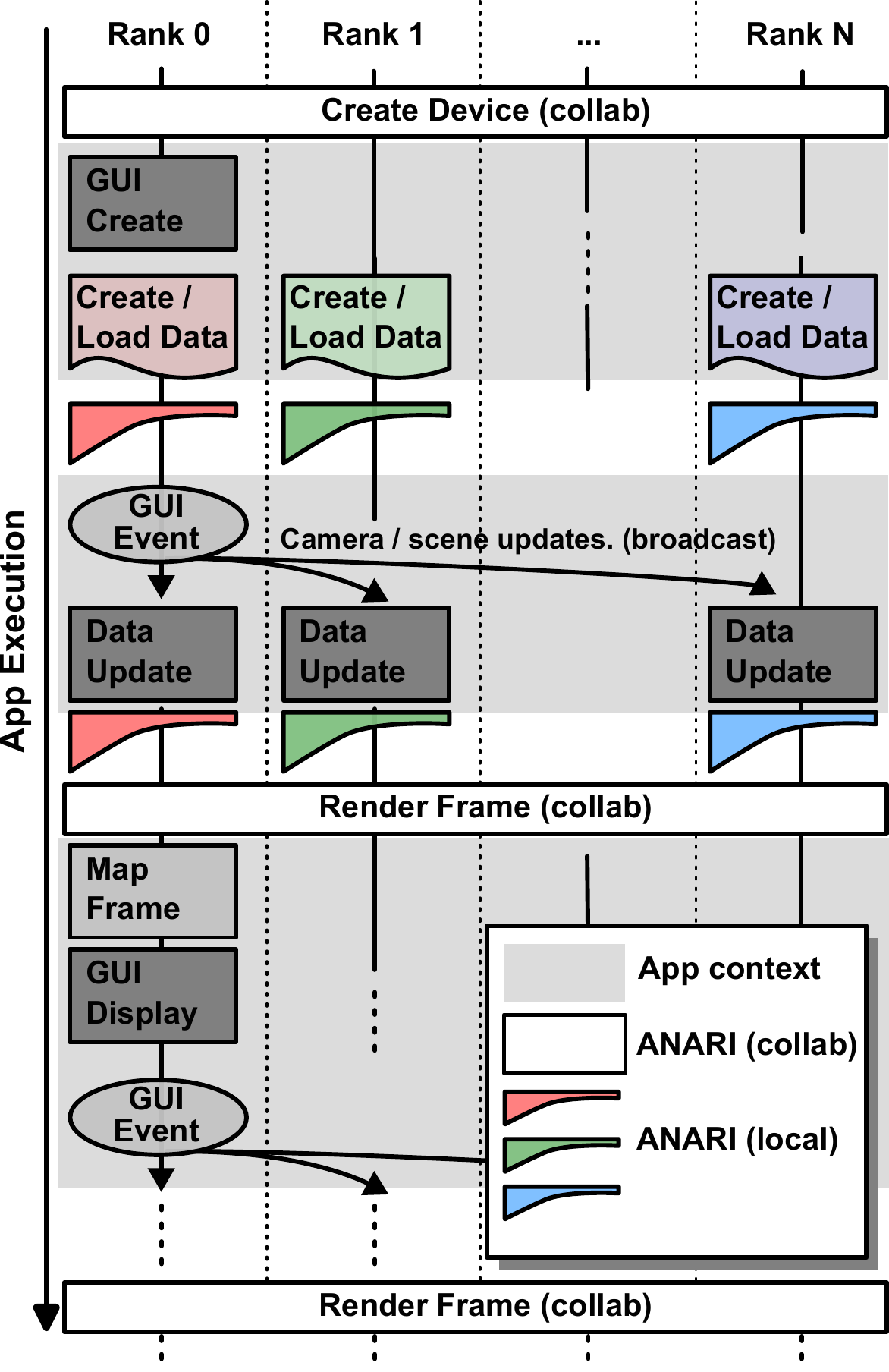}
 }
\end{center}
\vspace{-2em}
\caption{Illustration of how a typical data-parallel vis app would use our
  paradigm (also see \cref{sec:illustration}).
  \label{fig:illustration}
  \vspace{-1em}
}
\end{figure}

In \cref{fig:illustration} we illustrate exactly that workflow:
Upon startup all ranks would (collaboratively) create their DP-ANARI
device, then do whatever the app wants to do for data loading,
iso-surface extraction, etc. At some point each rank would create its
local ANARI world, and populate it with different ANARI
objects---these calls are not collaborative, so different ranks can do
as many of those---and whichever ones they want---without any other
ranks even being aware. Once all ranks are ready for rendering they
all call \code{anariRenderFrame()}, at which point they will
implicitly synchronize until the frame is done. Rank~0 can then map
the generated frame buffer, save or display it, and wait for UI events
or user input, at which point it will instruct its worker ranks to do
whatever scene updates are required, etc.

\section{Evaluation Challenges and Methodology}

What we have discussed so far is not a specific \emph{method}, nor a
\emph{system}. It cannot be evaluated via any one implementation, nor
via any one application using it, nor via any one or more use cases
thereof. Ultimately, the success of this proposed paradigm will depend
on whether---and to what degree---it will actually get adopted for
day-to-day data-parallel rendering, in tools such as ParaView or
VisIt.

Any such adoption is hard, because there is an inherent
chicken-and-egg problem that needs to be solved: applications will not
adopt any API or paradigm that there are no compelling device
implementations for, and for device implementers it makes little sense
to create such compelling implementations if there is no plausible
path for them to be adopted---nor is it easy to develop them if there
are no use cases to exercise them.

Eventually---and this is the purpose of the work described in this
paper---breaking this log-jam requires three \emph{simultaneous}
efforts: a) some example device implementation(s) that implement this
paradigm, and that end-user applications can actually target; b) some
reasonably complex applications that actually use this paradigm, that
use it in a way that is reasonably representative of how the eventual
applications-to-be will use it, and that device implementers can use
to develop, debug, and tune their implementations; and c) some
reasonably compelling proof-of-concept use cases that show that this
paradigm is actually worth adopting, and thus can serve as an
incentive for the other two parties to actually work towards this
goal.

\section{Example Realizations}

To serve as sample \emph{implementations} of our paradigm we
created two different ANARI devices that implement it. Both
devices implement the same paradigm, but are by no means directly
interchangeable: they do not offer the same feature set, nor will they
produce the same images for the same inputs. This is OK---different
applications have different needs, and different devices will always
offer different features. For our purposes, we intentionally chose two
opposite extremes of the spectrum: one relying on compositing, and one
performing true data-parallel path tracing.


\subsection{ANARI-Composite: Application-transparent compositing using an ANARI Pass-through Device} \label{sec:compositing-device}

The classical ANARI API---i.e., ANARI without the paradigms introduced
in this paper---has no concept of a ``data parallel world''. However,
that does not mean that applications \emph{using} ANARI could not do
data-parallel rendering of their own: such applications could use
ANARI to render data locally on each rank, and then rely on depth-
and/or alpha compositing to somehow composite the resulting
images. This approach has some obvious limitations in terms of what
effects can or cannot be rendered, but these limitations are not
specific to ANARI: they are not different from those when using any
other rendering back-end for the per-rank rendering.

Where these limitations are acceptable, one could use exactly
that same approach also \emph{within} an ANARI device to implement
the data-parallel ANARI paradigm we have introduced above. In such a device,
almost all functions would behave in exactly the same way as in any other local
device; namely, they would set up the scene to be rendered, and perform
rendering of a local frame buffer. The only thing this device would need
to modify is \code{anariRenderFrame()}, in which it would first perform its
local rendering, and then composite the results.  We decided to implement 
this approach, for which we need two main ingredients:
a means of performing local-node rendering, and a means of compositing.

\subsubsection{Compositing}

The standard way of implementing alpha and depth compositing is through
IceT~\cite{icet}; for our general use case this library is however too
restrictive: for alpha blending IceT requires that the application can provide
a fixed compositing order. With simple pass-through that compositing order is
however hard to determine as the input data would need to be classified into
opaque or transparent, the data convexly partitioned, etc., which would require
data specific knowledge the pass-through device cannot have.

To circumvent this we require that any fragment generated by the local device
for a pixel not only has an RGBA but also a depth component, which allows us
to instead rely on another compositing library we had readily available:
the \emph{deep compositing} (\code{deepComp}) library originally developed for 
another data-parallel rendering project~\cite{big-lander}. Using this
library an application (in this case, our ANARI compositing device)
can render possibly multiple RGBA-z \emph{fragments} for each pixel; the
compositor will then follow a \emph{parallel direct send}~\cite{pds,pds2} paradigm
to have all ranks exchange their fragments such that each rank gets all the 
fragments for some portion of all pixels. Each rank receives its pixels' fragments, 
and then, in a CUDA kernel, sorts and composites each pixel's fragments in proper
front-to-back alpha-composited order. The resulting composited pixels are
gathered at rank~0. For our purposes we only need a single RGBA-z fragment
per pixel (on each rank), but then no longer have to worry about which order
the different ranks generate these fragments.

\subsubsection{Rendering}

For the rendering, we can simply leave most of the heavy lifting to
ANARI itself, by using another \emph{existing} ANARI implementation for what we call
a ``pass-through'' device: except for very few calls that we describe below,
we simply pass all other calls through to this device, and can even let it
do the local rendering, as long as it is capable of computing both
color and depth buffers. We can even let the application choose which 
existing ANARI device to use, by intercepting the \code{anariLoadLibrary()}
call and dynamically loading this device as pass-through. 

To implement this compositing device, we can simply take every ANARI
API call and just pass it on to the pass-through device, except
for the following calls that we modify as follows:

\setlist[description,1]{leftmargin=1em,labelindent=0em}
\setlist{noitemsep}

\begin{description}
\item[\code{anariLoadLibrary()}:]
we first load our own data parallel device,
then on each rank also load the device requested by the application---and save
that as a pass-through device.
\item[\code{anariNewFrame()}:] we first pass this through to the
  pass-through device, and intercept the returned \code{ANARIFrame}
  handle for that rank. We then create our own---collaborative---frame
  object that first creates a new deep compositing context for that
  frame, and then also stores that rank's intercepted pass-through
  frame handle (through which we can later access that rank's local
  frame).
\item[\code{anariCommitParameters()}:] we pass this through to the pass-through device,
but also check if this commit affected a frame, and if so, whether it
resized that frame (and if so, resize that frame's compositing context). 
\item[\code{anariRenderFrame()}:] we pass this to the pass-through device to
  perform local rendering, then wait for that to finish, map the local rank's frame using the
  stored pass-through device frame handle, and perform compositing.
  Compositing in \code{deepComp} requires collective MPI calls, but in
  our data-parallel ANARI paradigm \code{anariRenderFrame} is collective,
  so this poses to problem.
\item[\code{anariMapFrame()}:] this is the one call we do not pass
  through at all, since local frames have already been read and
  composited in \code{anariRenderFrame}. We simply retrieve the composited
  image from the compositing context, and return this.
\end{description}

\noindent
One advantage of this approach is that it is easy to implement: given
an existing compositing library we implemented a working proof of
concept with very little effort, in less than a day.
%

What is particularly useful is that because our compositing device
itself works by issuing ANARI calls to the pass-through device it can
actually use \emph{any} other exiting ANARI device for the actual
rendering, without having to know which. This makes this approach
useful as an easy ``fall-back'' mechanism of using any other---not yet
natively data parallel---ANARI device in a data parallel context.

The downside to this approach is that it is intrinsically
limited to what compositing can or cannot do. Using the \code{deepComp}
library means we can avoid some of the specific limitations of
IceT---in particular, we do not need to specify a fixed compositing
order---but it still relies on compositing, and thus will never be
able to produce guaranteed-correct shadows or path tracing for data
parallel content.


\subsection{Barney and (B)ANARI} \label{sec:barney}

Barney is a new---and still under development---project for
data-parallel path tracing on multi-node and multi-GPU hardware. For
rendering, Barney relies on ray forwarding similar to what is done by
the recently published Brix~\cite{brix} and RQS~\cite{rqs} papers,
where rays are sent to the node(s) that may have geometry that may
intersect a given ray---and where each ray will always find its
respectively closest intersection no matter which rank the ray was
spawned on, or which rank holds that respective geometry.

Unlike ANARI, Barney was built with parallel rendering---and in
particular, data parallel rendering---in mind from its very
inception. In addition to the relatively simple mode we described
above---where each rank has exactly one part of the data to be
rendered---Barney also offers various additional modes such as, for
example, data-replicated rendering, islands-parallel rendering,
non-MPI multi-GPU rendering, additional multi-GPU data-parallel
rendering within a given rank, etc. Despite this bigger set of
functionalities, Barney follows the same general paradigm described
above: it has the concept of a data-parallel world, different ranks
can independently specify different pieces of this world, and render
operations need to be synchronous.

\subsubsection{BANARI}

Barney is not \emph{exclusively} built for ANARI, but targets the same
end-user applications, and thus supports similar functionality:
it supports both surface and volume types, and in particular also
supports the more sci-vis oriented data types of cylinders and
spheres for surface data, or unstructured mesh and AMR data for volume
data. 

With all these pieces in place, implementing a data-parallel ANARI
device was relatively straightforward: mostly this required to
implement the various ANARI API functions to properly read the render
data passed through this API, and passing it on to its matching data
types, where applicable. As with many other ANARI implementations,
the Barney ANARI device---or \code{BANARI}, for short---implements
only a subset of the full set of ANARI's different data types, and
simply ignores all others.


\subsubsection{Local vs.\ Global Rendering}

Barney is, by nature, designed for data-parallel rendering. However,
to also be accessible to non-data parallel applications it can also be
built without MPI support, in which case it simply performs local,
data-replicated multi-GPU rendering.

If built this way, the BANARI device still implements the ANARI API
(just without out data-parallel semantics), and we can thus also use
this as a pass-through device for the ANARI compositing device
described in \cref{sec:compositing-device}.


\section{Example Integrations}

Whereas the previous section showed that it is possible (and in fact,
not all too hard) to write devices that implement our paradigm, in
this section we look at the reverse problem of integrating such an API
in an application that wants to use data-parallel ANARI.

\subsection{Minimal, Proof-of-Concept Applications}

We are ultimately most interested in how hard it is to integrate our
paradigm into actual end-user applications like VisIt or ParaView, or
widely used frameworks such as VTK. However, we could not evaluate that
until some implementation(s) existed, but neither could we have
developed the aforementioned back-ends without applications to exercise
it. To break this chicken-and-egg problem we decided to defer
integration into actual end-user tools to the end, and in the
meanwhile, relied on developing both back-ends and several different
front-ends in parallel.

\subsubsection{OSPRay and TSD Mini-Apps}

As one proof-of-concept we started with several existing mini-apps from
the OSPRay project. The semantics of our data parallel ANARI and those
of data parallel OSPRay are (intentionally) quite similar; the API
calls and function names differ, and so do some low-level concepts,
but the application flow is similar. To prove the generality of data
parallel ANARI we took the data parallel sample apps coming with
OSPRay and ported them to our semantics---where possible
line-by-line. This allowed us to exercise our semantics on some very
simple data parallel apps outside our own software ecosystem.

This provided an early proof of concept, but does not come with much
interesting data to test with. To get more realistic inputs, our next
step was to take an existing single-rank ANARI viewer (\code{TSD},
from \code{VisRTX}~\cite{visrtx}), and prototypically extended that to
use MPI parallelism: all ranks load different parts of the input, then
rank~0 runs the existing viewer, and broadcasts UI updates and render
requests to the worker ranks; workers wait for such broadcasts, then
perform ANARI scene updates and call \code{anariRenderFrame}. This all
naturally follows our paradigm, meaning all the effort in this
proof-of-concept was in changing this viewer to be MPI-parallel in
itself---with no extra work for our paradigm at all.




\begin{figure*}[t!]
  \resizebox{1.00\textwidth}{!}{
    \includegraphics[height=2.5cm]{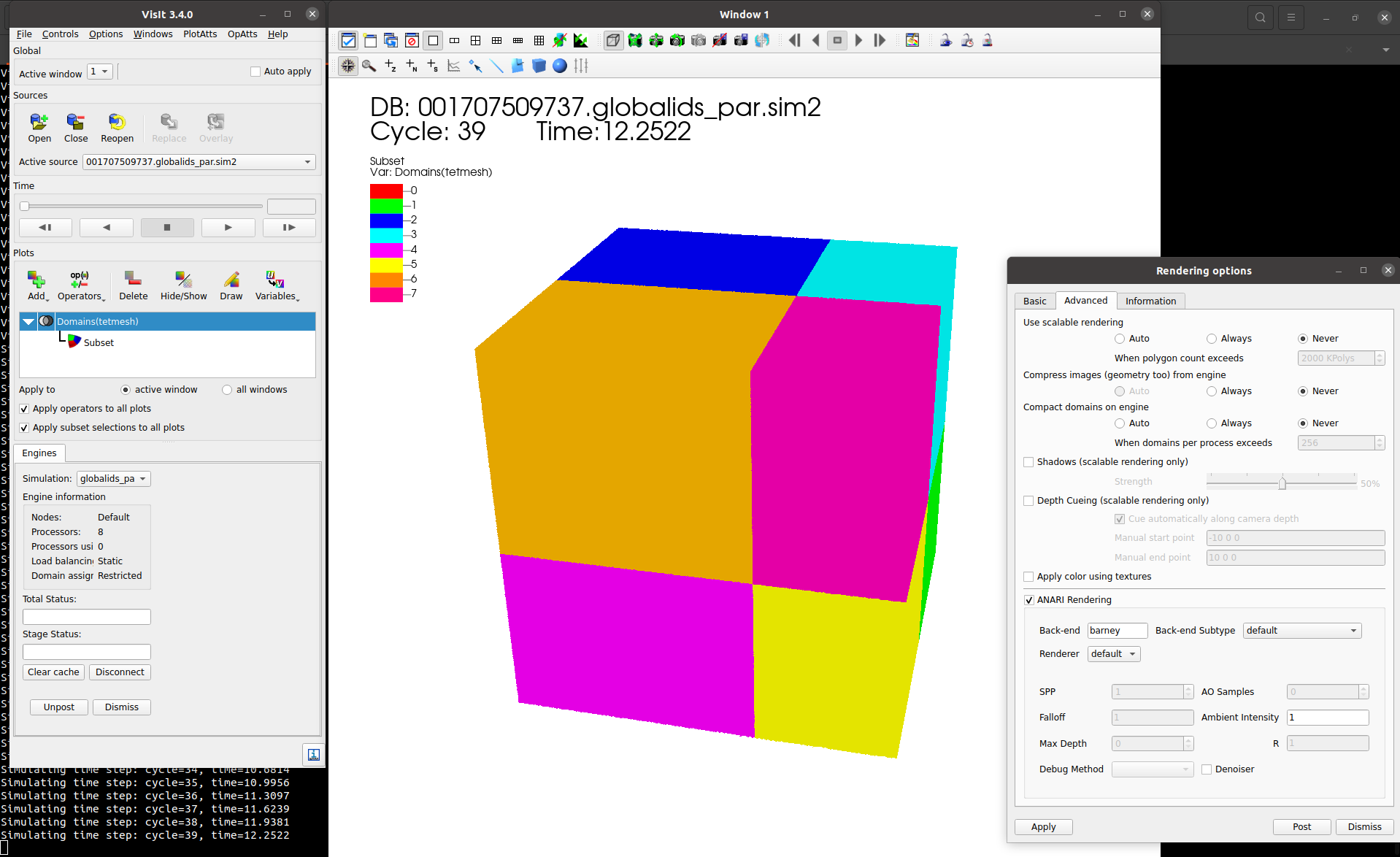}
    \includegraphics[height=2.5cm]{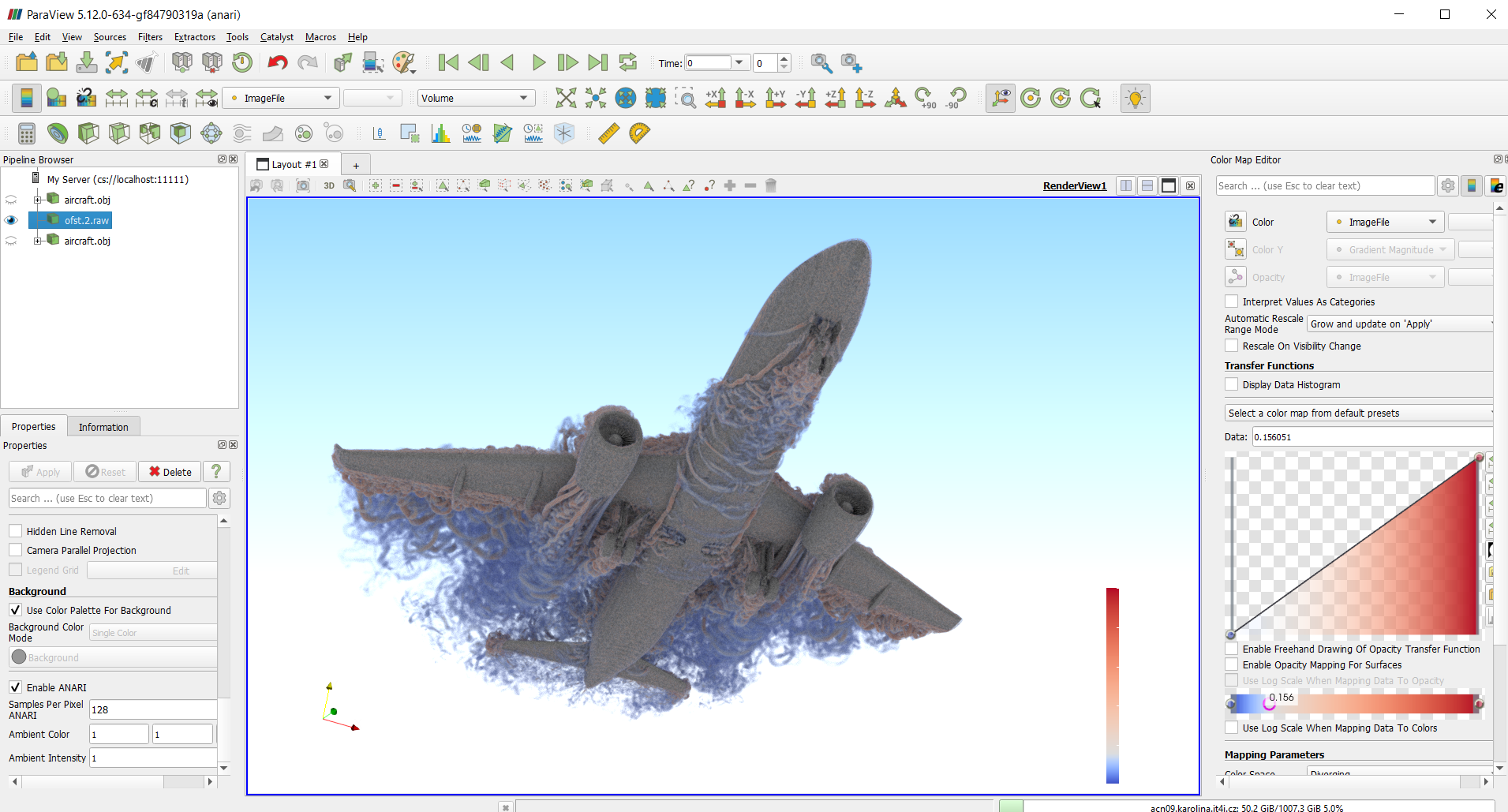}
    \includegraphics[height=2.5cm]{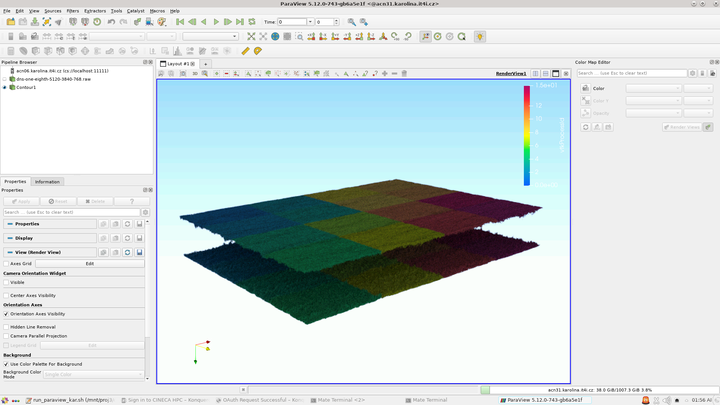}
    \includegraphics[height=2.5cm]{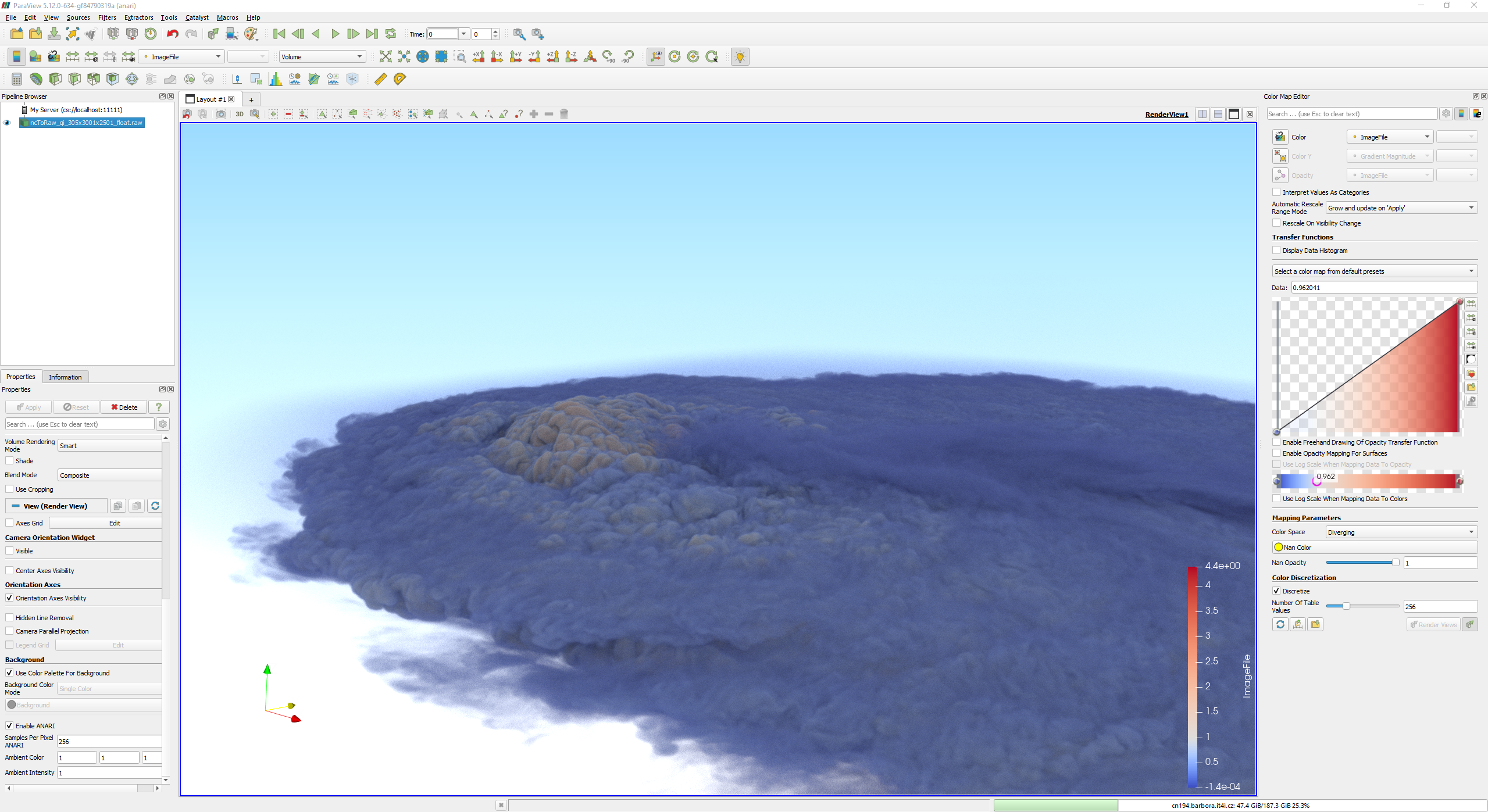}
  }\\[-4ex]
  \caption{Screenshots from our prototypical
    integrations of our DP-ANARI paradigm into VisIt/Libsim and
    ParaView. 
    a)~VisIt/Libsim data set generated by a mini-simulation used to
    test VisIt's in-situ capabilities, running on 8 ranks and using
    our data-parallel ANARI integration (see \cref{sec:libsim}).
    b)~ParaView rendering the $1,225\times 580\times 2,000\times$\code{float}
    \emph{aircraft} model with Q-criterion vortex field (see \cref{sec:paraview}).
    This case was rendered with BANARI on 4 ranks of 1 GPU each.
    c)~ParaView rendering an iso-surface extracted from a $2\times$ reduced version of the
    DNS data set.
    d)~ParaView rendering the $2,501\times 3,001\times 305\times$\code{float}
    \emph{thunderstorm} data set, in this case rendered with BANARI
    on 4 ranks of 1 GPU each (using Barney's volumetric path tracing
    mode).
  \label{fig:visit-and-paraview}
  \label{fig:visit}
  \vspace*{-1em}
  }
\end{figure*}

\subsubsection{HayStack, and HANARI}

\def\haystack{\code{HayStack}}

In order to have a somewhat more challenging use case we also
took Barney's original \haystack viewer, and prototypically ported
(parts of) that over to ANARI. \haystack was originally developed for
Barney even before Barney's ANARI device was developed.

\haystack is what we consider a developer-centric minimal GUI
app with the minimum at what it takes to manipulate cameras, edit
transfer functions, etc. While \haystack is intentionally minimalistic
when it comes to the user interface, it was from the beginning
designed to be able to stress-test Barney to the end of its abilities,
and to be as realistic a mock-up of what a real application like
ParaView would be as it possibly could. In particular, \haystack
supports importers for many different data types including structured
and unstructured volume data; triangle meshes, spheres, cylinders, and even
production-style data with instances and textures; it contains
facilities for data-parallel loading, offline and on-the-fly
partitioning (including both object-space and spatial partitions),
where desired, different data load balancing schemes (to simulate
different ways of how an application might assign geometry to
different nodes).

While originally \emph{not} under ANARI, this wealth of different data
configurations made it an attractive candidate to also add an
ANARI-based render pass. As with the mini-apps, virtually all the work
required in doing so is the traditional ANARI calls to create and
provision each rank's geometry, with no special effort to meet our
DP-ANARI requirements whatsoever. We observe, however, that this ANARI
path currently only supports some of \haystack's data types, and even
then does not necessarily produce exactly the same images. Usually,
this is because some of Barney's data formats and material types are
slightly different from ANARI's. Nevertheless, the key insight from
this exercise matched exactly what we saw for the significantly
simpler mini-apps: almost all the effort required for using a DP-ANARI
device lies in how a given rank specifies its ANARI---which is exactly
the same as it would have been for per-rank ANARI rendering---with
virtually no extra effort required for the data-parallel portion.


\subsubsection{Data Parallel VTK Mini-App} \label{sec:vtk}
\label{sec:vtk-miniapp}

Being one of the most popular sci-vis frameworks, interoperability with the
Visualization Toolkit (VTK)~\cite{vtkBook} was an obvious choice for us. In
preparation for further integration with ParaView~\cite{paraview} we started
with a simple proof of concept application using VTK internal classes only.

To add a new rendering subsystem to VTK, a custom ``pass'' needs to
be registered with the \code{vtkRenderer}. VTK already has builtin support
for rendering with ANARI via the \code{Rendering/ANARI} module, which
co-exists with other rendering modules (e.g., for core OpenGL, or
OSPRay), and which we obviously chose to build on. 

VTK itself has no concept of any data parallel rendering; any data
parallel loading and rendering has to happen on the application
side. To do this we started by implementing a simple MPI-parallel
app in which each ranks loads its respective geometry, assuming
pre-partitioned data on disk. For the user interface we created a
simple VTK GUI app using the \code{vtkRenderWindow} class, which
implements a platform-specific event loop utilizing GLX, WGL, or
similar depending on the target platform, and calls the
\code{Render()} function of the \code{vtkRenderer} object attached
when redraw events occur. Simply attaching a \code{vtkAnariPass}
object to the latter we achieve serial rendering.

The actual GUI window only runs on rank~0, but to make our paradigm
work, the workers also need to run the same VTK render passes, in
synchronous mode. We solve that by having each worker create an
off-screen window of the same window class that the display rank uses:
this creates the same pipeline as on the display rank, but obviously
does not get any of that rank's UI events. We solve this by having the
display rank perform MPI broadcasts for events that affect global data,
such as resize, camera changes, or render. Workers then implement a custom
event loop in which they listen for such requests from the display rank,
upon which they then issue the corresponding events.

One caveat with the standard \code{vtkRenderWindow}, which was not
implemented with data parallelism in mind, is that events that require
lockstep processing can be triggered unexpectedly; on X11 for example,
a resize event \emph{plus} redraw event only gets triggered when the
window size increases. To guarantee lockstep execution, we
implemented a pass-through extension via a C++ class inheriting from
\code{vtkAnariPass} that intercepts these events and communicates with
the worker's event loop using a simple custom communication protocol
on top of MPI.

In retrospect, most of the challenges of this exercise turned out to
be related to VTK---and in particular its event system---not having
any concept of other clients it might have to synchronize with, and
wanting to issue supposedly-synchronous ANARI calls at
seemingly-random UI events. Somewhat unexpectedly, this meant that the
integration into nominally much more complicated applications (see
upcoming sections) turned out to be actually \emph{easier} than what
was originally intended as a warm-up exercise for these applications.

\subsection{Prototypical Real Vis-App Integrations}

While mini-apps and \haystack were invaluable from a developer's point
of view, the eventual goal for any data-parallel ANARI effort must
necessarily lie with actual ``real'' end-user vis apps such as
ParaView, VisIt, or various in-situ frameworks.

\begin{figure*}[ht]
  \centering
  \resizebox{1.00\textwidth}{!}{
    \stackunder[8pt]{\includegraphics[height=33mm]
      {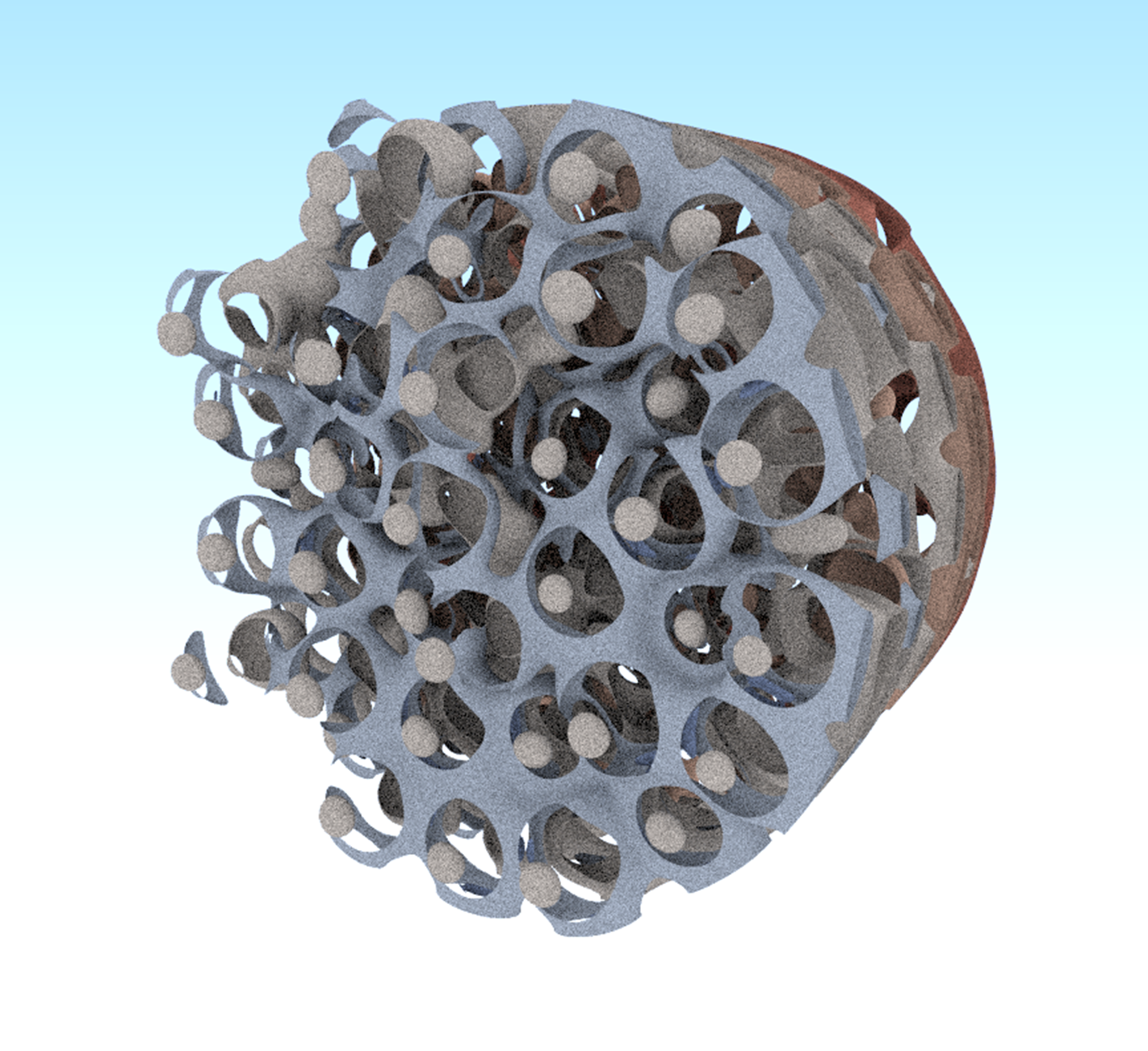}
      }{NekRS: \code{pb146} pressure}
    \stackunder[8pt]{\includegraphics[height=33mm]
      {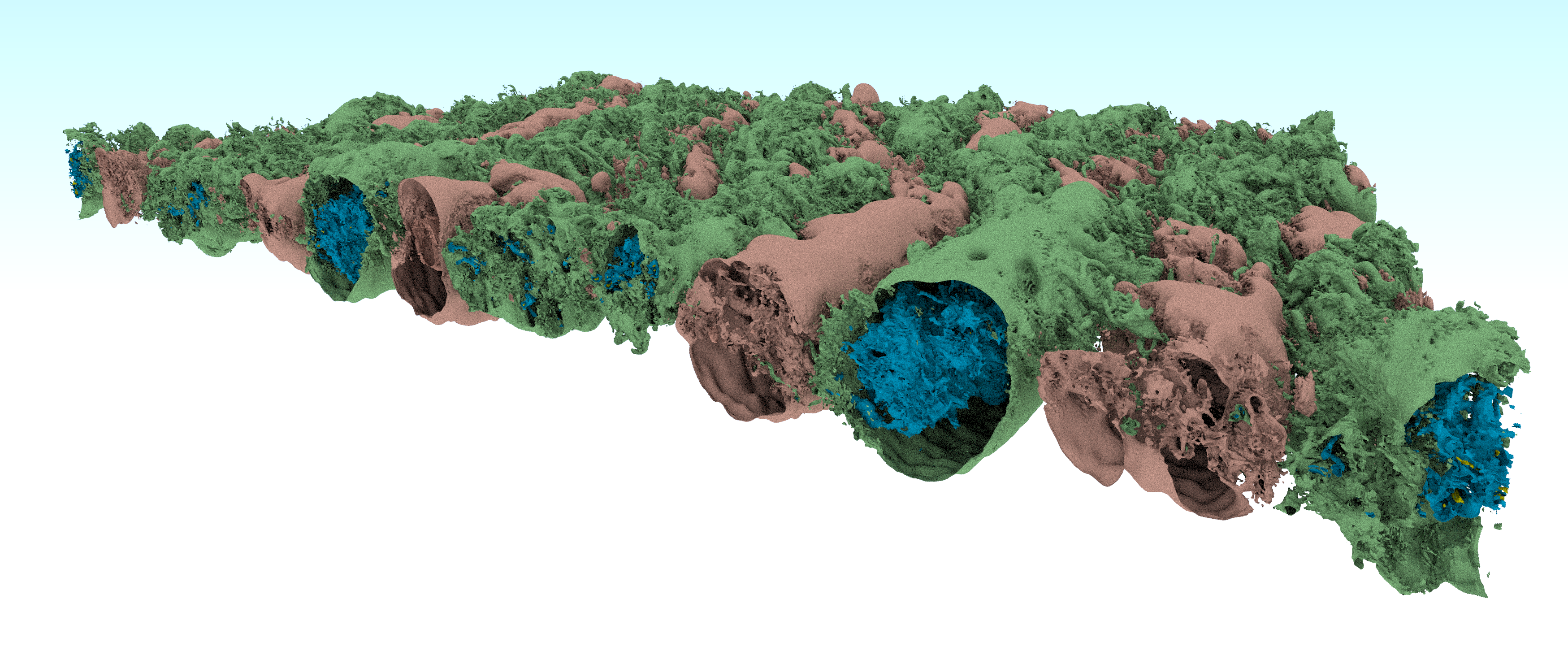}
      }{S3D: \code{ammonia} pressure (iso-surface)}
    \stackunder[8pt]{\includegraphics[height=33mm]
      {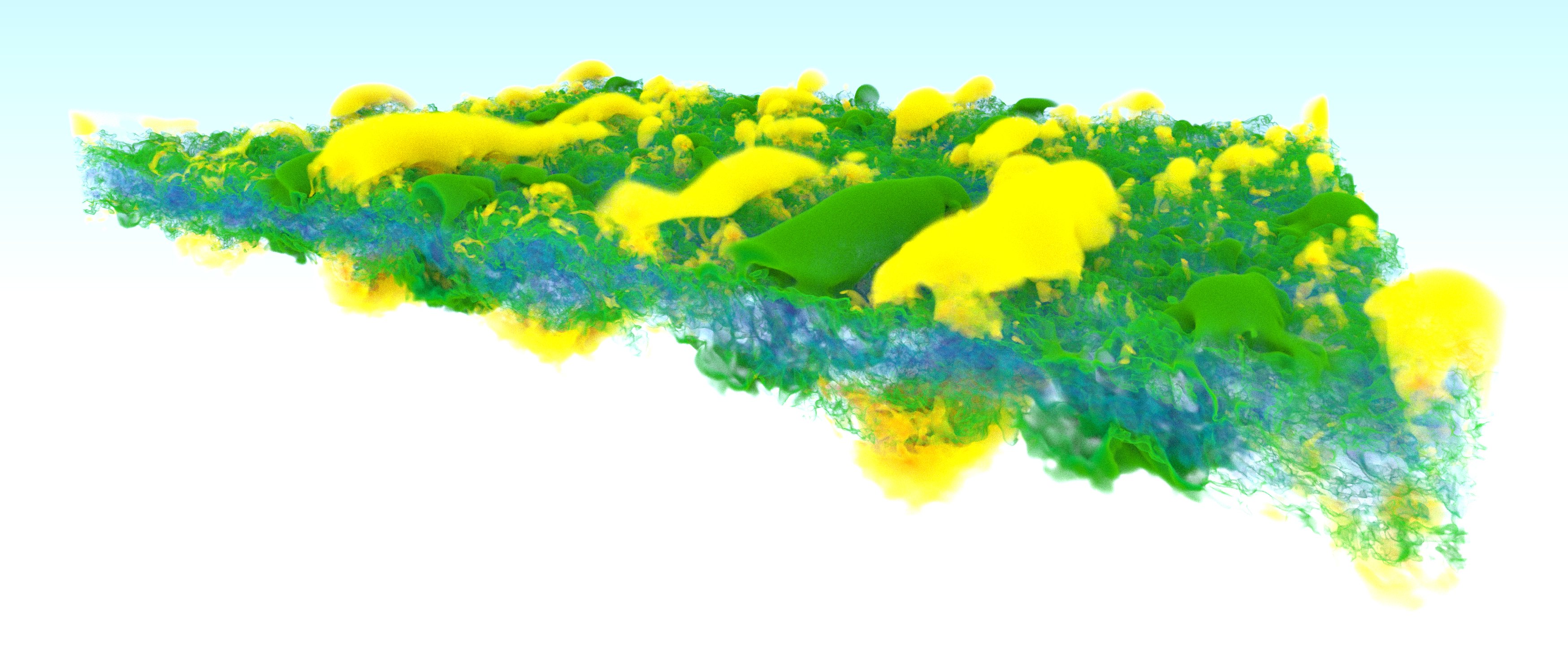}
      }{S3D: \code{ammonia}, velocity magnitude (volume rendered)}
  }\\[-1ex]
  \caption{
    Some images from our ASCENT integration, used on two different HPC simulation codes and data sets (rendered using BANARI).
    \label{fig:ascent}\vspace*{-1em}
  }
\end{figure*}


\subsubsection{ParaView}
\label{sec:paraview}

To implement our DP-ANARI semantics into ParaView~\cite{paraview} we
start with the proof of concept app described in \cref{sec:vtk}.  In contrast
to VTK, ParaView supports full data parallel rendering via a client/server
architecture. Synchronization issues such as the ones encountered with
\code{vtkRenderWindow} in our test app hence are not an issue. To run
ParaView in data-parallel mode, we start the \code{pvserver} app with
as many MPI ranks as we have workers, and connect to that using
the ParaView client. In that setup, the ParaView client does not participate in
the MPI-parallel rendering, but will receive remote-rendered, compressed
images from the dedicated MPI worker rank that eventually holds all the pixels.

Although exposing ANARI-based rendering in ParaView is straightforward
(using similar routines as for OSPRay), this extension has not yet found
its way into the upstream repository; our work builds off a custom
implementation comprising a handful of ParaView classes
to expose VTK's \code{Rendering/ANARI} in the GUI.

ParaView realizes compositing via IceT, but for our paradigm it does
not make sense that the application should do that. One option would have
been to take this component out of ParaView, but this would have
required non-trivial changes to ParaView.  Instead, we simply have
rank~0 pass the already-finalized image to IceT, while all other ranks
report an empty frame. When IceT ``composits'' these frames it will
simply end up getting what we reported on rank~0; it will have
performed some un-necessary computations, but since compositing is not
a bottleneck, this is acceptable to us. Using this approach all our paradigm's
demands are already met: rendering and resize are synchronous (IceT had this
requirement, too), and all per-rank rendering is done through ANARI already.

The only snag we hit with this approach is that ParaView currently
uses an optimization by communicating the local scene bounds to IceT
on the worker ranks; this allows IceT to determine that certain pixels
will not have any (directly) visible geometry, and thus can be
excluded from compositing. This optimization obviously does not work
for our approach where rank~0 always reports all pixels. We currently
disable this optimization, in which case our approach works as
expected. We also observe that this problem only occurs because we
decided to leave IceT enabled in ParaView; the moment ParaView were to
fully adopt data parallel ANARI it would be cleaner to disable
composing on the app layer, anyway, at which point this problem would
no longer exist.

\subsubsection{Ascent}

To also experiment with an in-situ scenario, we also integrated our
data parallel ANARI approach into Ascent, which is an increasingly
popular, lightweight in-situ visualization and analysis
infrastructure~\cite{ascent}.  Ascent uses VTK-m, not VTK, so the VTK
work from \cref{sec:vtk-miniapp} did not apply. We therefore
decided to base our integration directly on the ANARI level, which we
then did by implementing new ``extracts'' in Ascent (i.e., to follow
using a customized renderer type in Ascent's terminology).

In this framework, meeting our DP-ANARI requirements
was trivial. Consequently, virtually all the work in this
integration was in issuing the ANARI calls to create each rank's local
scene data, which is identical to what would have been needed for a
local per-rank ANARI back-end, too. In fact, using DP-ANARI meant that
we \emph{only} had to do the ANARI geometry calls, and not worry about
any compositing of the results, which in DP-ANARI is not required.

Unlike the previously mentioned mini-apps, this integration was not
just done for the sake of evaluation---but as an actual means of
enabling high-fidelity data parallel rendering for actual state of the
art simulations. In \cref{fig:ascent} we show two examples of
this: \texttt{NekRS}, a GPU-accelerated spectral element Navier-Stokes
Solver for incompressible turbulent flows employing an unstructured
hexahedral mesh~\cite{fischer2022nekrs}; and \texttt{S3D}, a scalable
direct numerical solver for reactive and compressible flows, based on
a rectilinear mesh~\cite{hawkes2007scalar,chen2009terascale}.

\subsubsection{VisIt/LibSim}
\label{sec:libsim}

As a final proof of concept we are also working on an integration into
\code{libsim}~\cite{libsim}. Libsim is an infrastructure for in-situ
visualization using VisIt~\cite{visit}, so some sense this effort is
closer to the Ascent effort than it is to our ParaView
integration. However, libsim connects to VisIt for visualization which
uses VTK for rendering, so the actual steps to make it work are
essentially identical to what we have described above for the VTK
mini-app, and works exactly the same.

For the initial testing we used the globalids
mini-simulation~\cite{globalids} and are thus limited to the
relatively simple test geometry produced by this libsim
mini-simulation. This is obviously not representative for the kind of
volume or geometry data that a real libsim session would
generate. However, what geometry and volumes libsim generates on a
given rank should only affect what local per-rank operations the
underlying ANARI VTK renderer would need to perform. While this
clearly needs more testing and profiling, we believe that the key
concepts are already in place. A screenshot of this effort is shown in
Figure~\ref{fig:visit} (in this case, rendering through BANARI)



\section{A Case for Adopting this Paradigm}

In the last two sections, we have shown that it is reasonably easy to
\emph{realize} our DP-ANARI paradigm in back-ends, and that it is
similarly easy to \emph{integrate} it into new or existing
data-parallel applications. Together, these two arguments show
that our proposed paradigm is \emph{realistic} in the sense that if
one is seriously interested in standardizing towards an API for
data-parallel ray traced rendering, then this paradigm could indeed
work. In this section, we are making an argument as to \emph{why} any
given application should care.

\removed{We are not going to re-argue the case for an ANARI standard (and what
applications have to gain from adopting it)---this argument has been
made before, and the wide-spread adoption of (single-rank) ANARI is
evindence that it has been heard. Instead, we focus on why---in
particular in sci-vis---we have to go beyond single-rank ANARI, and
instead start considering data parallel ANARI.}

\subsection{Impact on Existing, Classic-ANARI Apps}
\label{sec:discussion-single-rank}

\removed{The first argument we are making is that the way our paradigm is
purely additive in that it does not take \emph{anything} away from
existing state of the art. For example, consider an existing
single-process (i.e., not data parallel) application that already has a
means of using classical ANARI---such as, for example, TSD or Blender (cf. Fig. ??).
The way our data-parallel ANARI works is that for a single rank or
process is exactly identical to classical ANARI, because classical
ANARI automatically fulfills the rules laid out above: each rank can
have its own data (there is only one, and it has all the data),
anariRenderFrame is synchronous across all ranks (one rank is always
synchronous), etc. In reverse, this means that any exising ANARI apps
can simply use any DP-ANARI back-end without any change whatsoever.}

\removed{Similarly, consider an existing data-parallel app that already uses
ANARI for local, per-rank rendering, even if it still does its own
compositing---such as, for example, mainline ParaView (the existing
one, without our modifications). This, too, could use any DP-ANARI
capable device for per-rank rendering, without even noticing---it
would not yet benefit from our paradigm's advantages, but would also
see no negative effects whatsoever. In particular, it could use our
compositing device and pass through to whatever device it used before,
and would see no difference whatsoever.}

\begin{figure}[ht]
  \resizebox{1.00\columnwidth}{!}{
    \includegraphics[height=2.5cm]{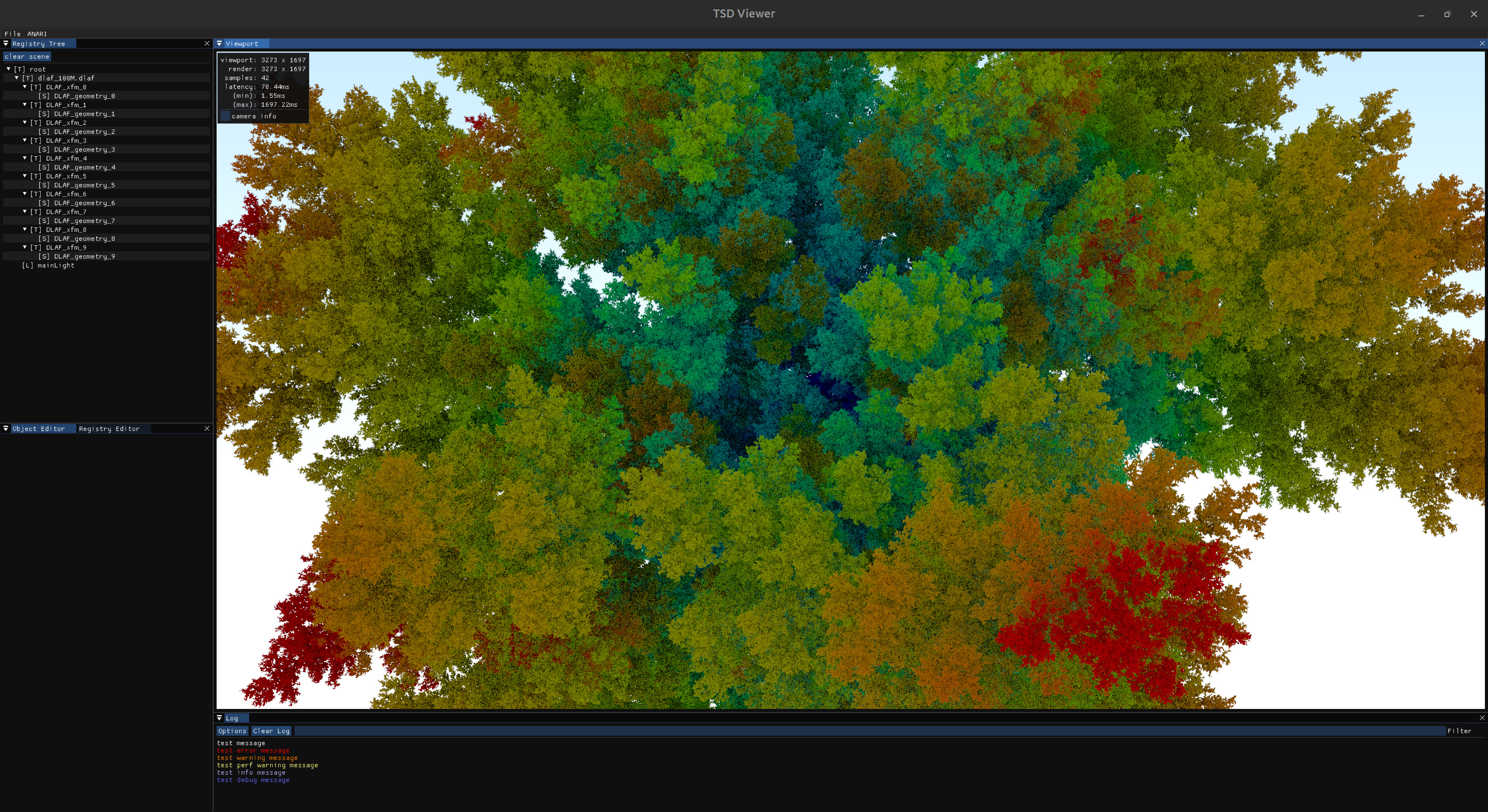}
    \includegraphics[height=2.5cm]{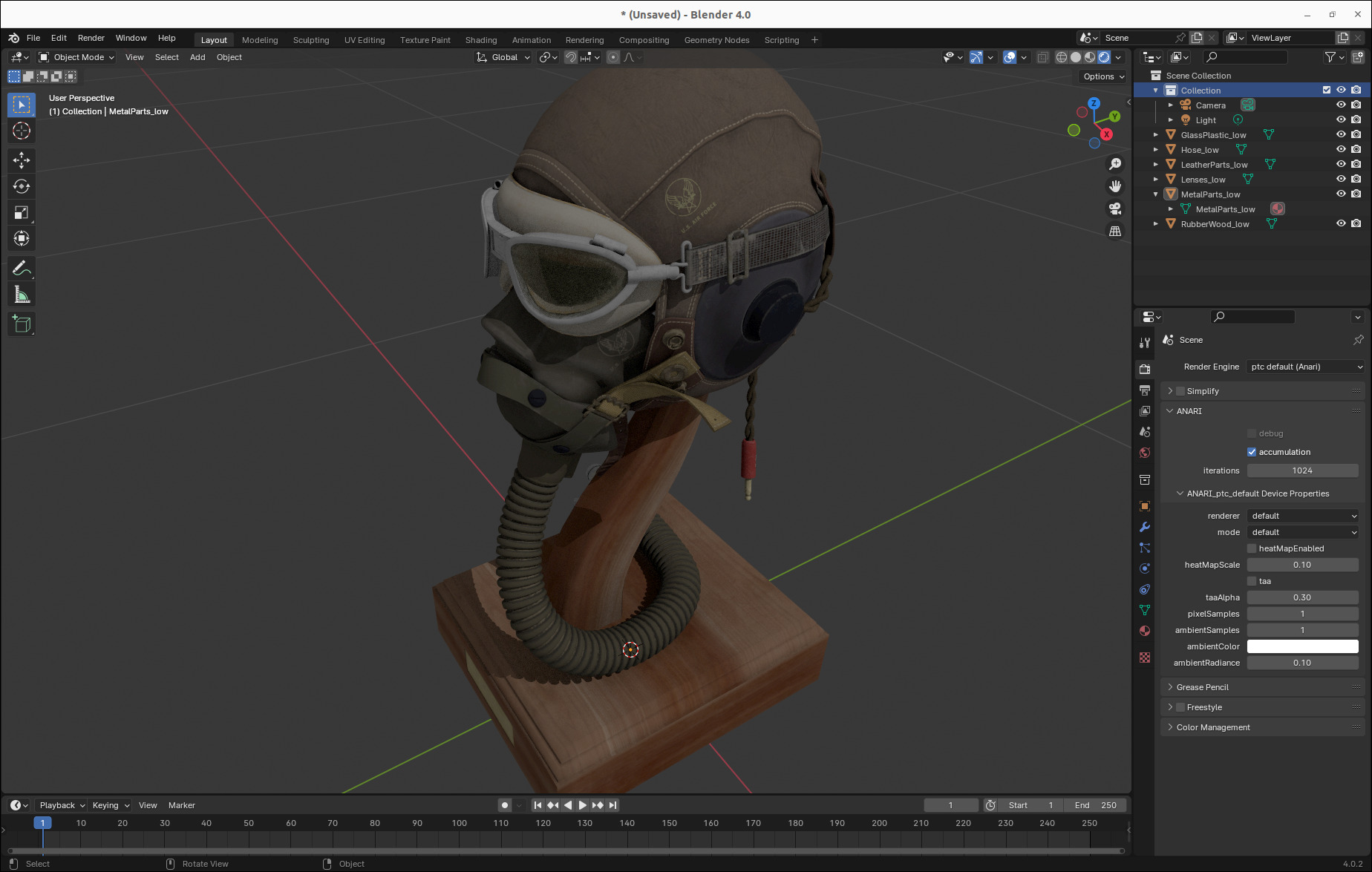}
  }\\[-4ex]
  \caption{Examples of our paradigm on traditional---i.e., \emph{not}
    data parallel---applications: a) a stand alone ANARI-based
    viewer (in this case, using BANARI). b)~\code{blender}, using
    \code{ptc/visionaray}. Though our paradigm does enable data-parallel path tracing,
    it does not change anything for traditional, non-parallel apps,
    which can therefore use our DP-enabled devices without any changes.
    \label{fig:not-data-parallel}
    \vspace*{-1em}
    }
\end{figure}

\added{Our first observation is that our paradigm is purely additive in that it
does not take \emph{anything} away from existing state of the art. A single-process app
already using ANARI (e.g,, TSD or Blender (cf.\ \cref{fig:not-data-parallel})
will use ANARI in the exact same way, which will have the exact same performance implications
as it did before. Similarly, an existing data-parallel app that however already
uses ANARI for local, per-rank rendering (e.g., mainline ParaView without
our modifications) would also see no negative effects whatsoever when using
the DP-ANARI paradigm and whatever means it uses to render images in parallel.}

Let us now consider this same app to follow the (simple) steps
outlined above (\cref{sec:paraview}) to actually enable that
device's data-parallel capabilities, and follow our paradigm. Assuming
it simply used our compositing device (passing through to whatever it
used before) it still would not see any difference whatsoever. At this
point the app could already get rid of its own compositing layer, and
still not ``lose'' anything it could have done before. Though the app
would still not have seen any benefit, it also would not have lost
anything by adopting our paradigm.

\subsection{Benefits of Adopting a DP-ANARI Paradigm}

\removed{Having just argued that an application has no reason \emph{not} to
adopt our DP-ANARI paradigm, the argument \emph{for} adopting it is
simply the same as the argument for ray- and path-tracing in vis,
which has been made numerous times in recent years (see,
e.g., [?]): Ray tracing allows for advanced
shading effects such as shadows, indirect illumination, better volume
rendering and volumetric path tracing, etc---but for either one of
those effects a ray tracer \emph{must} have some information about all
data in the scene, no matter which rank has it. Without this
information a data-parallel app could still use compositing on locally
ANARI-rendered images, but could only do so without these shading
effects, even if each rank's image is locally using a ray tracer. To
illustrate this, in Fig ?? we have
created an intentionally simple test case, and render that once with
true data parallel path tracing, and once with compositing over
per-rank but not globally ray traced images. To make this as
comparable as possible we use the same path tracer (Barney) in both
cases: for the true data parallel case we use full BANARI; for the
reference case we built BANARI without MPI support (so it only
performs single-rank rendering), then use that as a pass-through
device for our compositing device. As can be expected, shadows are
largely missing from the reference image because shadow rays on a
given rank cannot ``see'' the shadow casters that other ranks have.}

\begin{figure}[ht]
\begin{center}
  \includegraphics[width=.32\columnwidth]{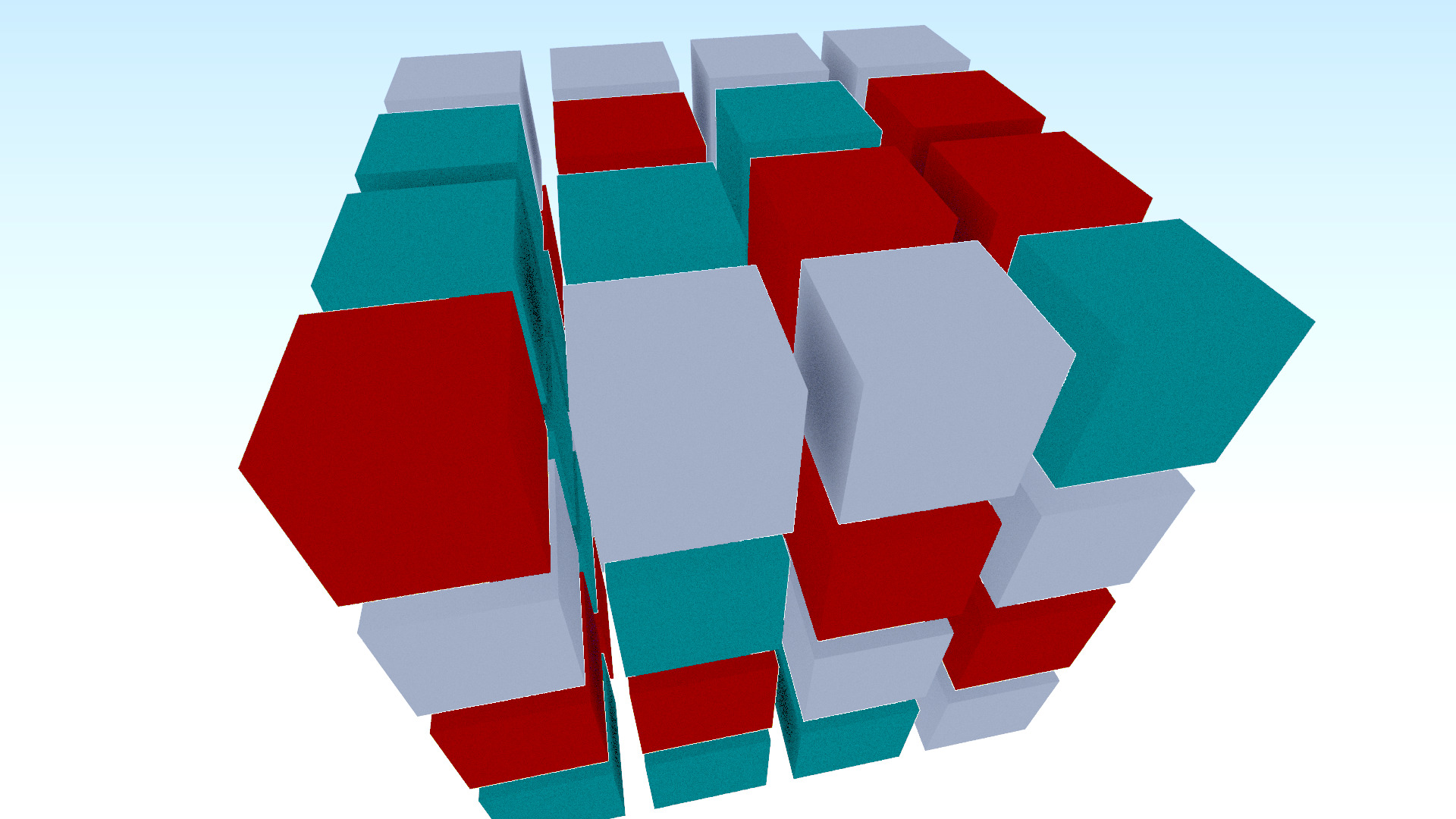}
  \includegraphics[width=.32\columnwidth]{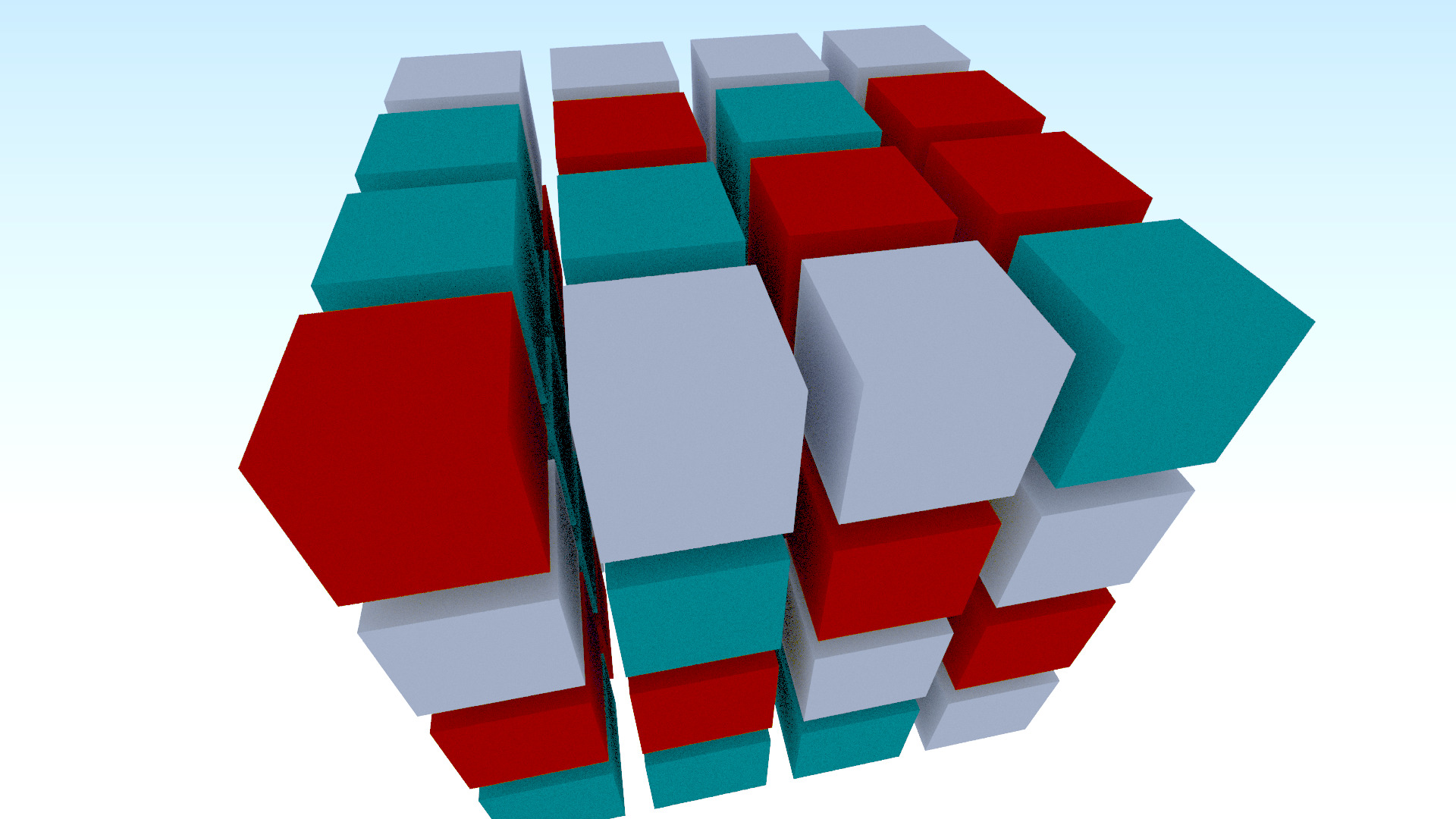}
  \includegraphics[width=.32\columnwidth]{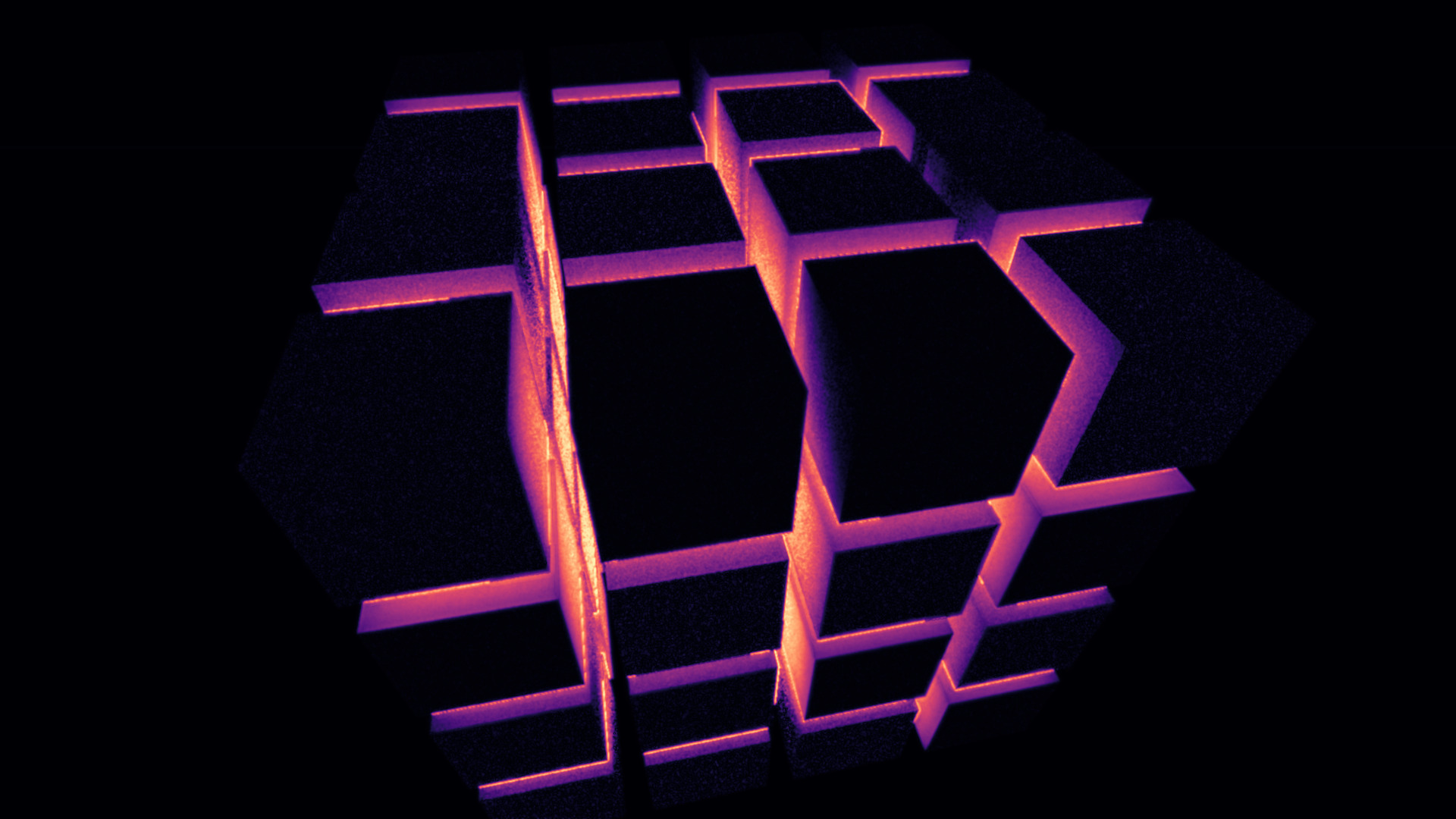}\\
  \includegraphics[width=.32\columnwidth]{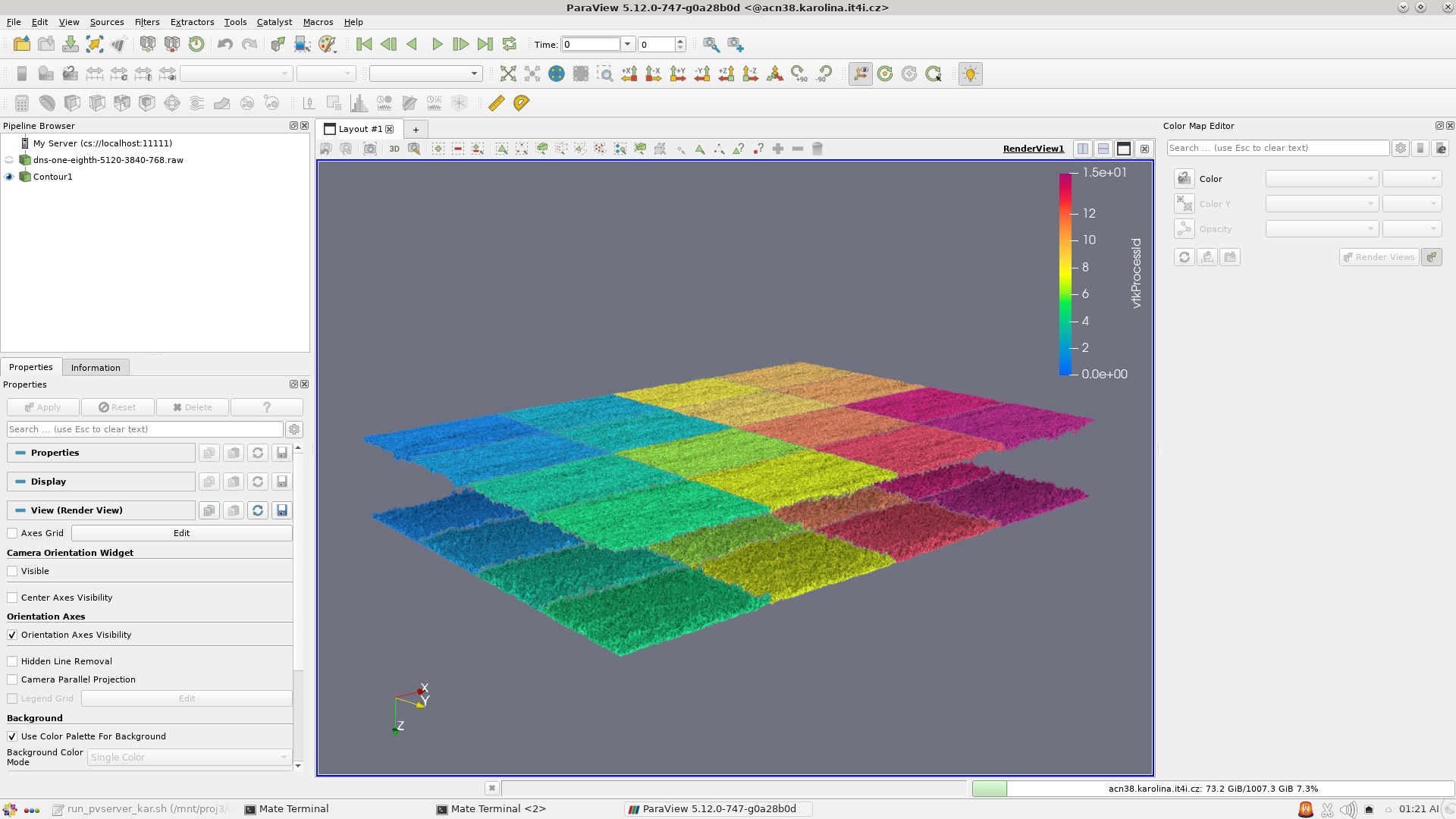}
  \includegraphics[width=.32\columnwidth]{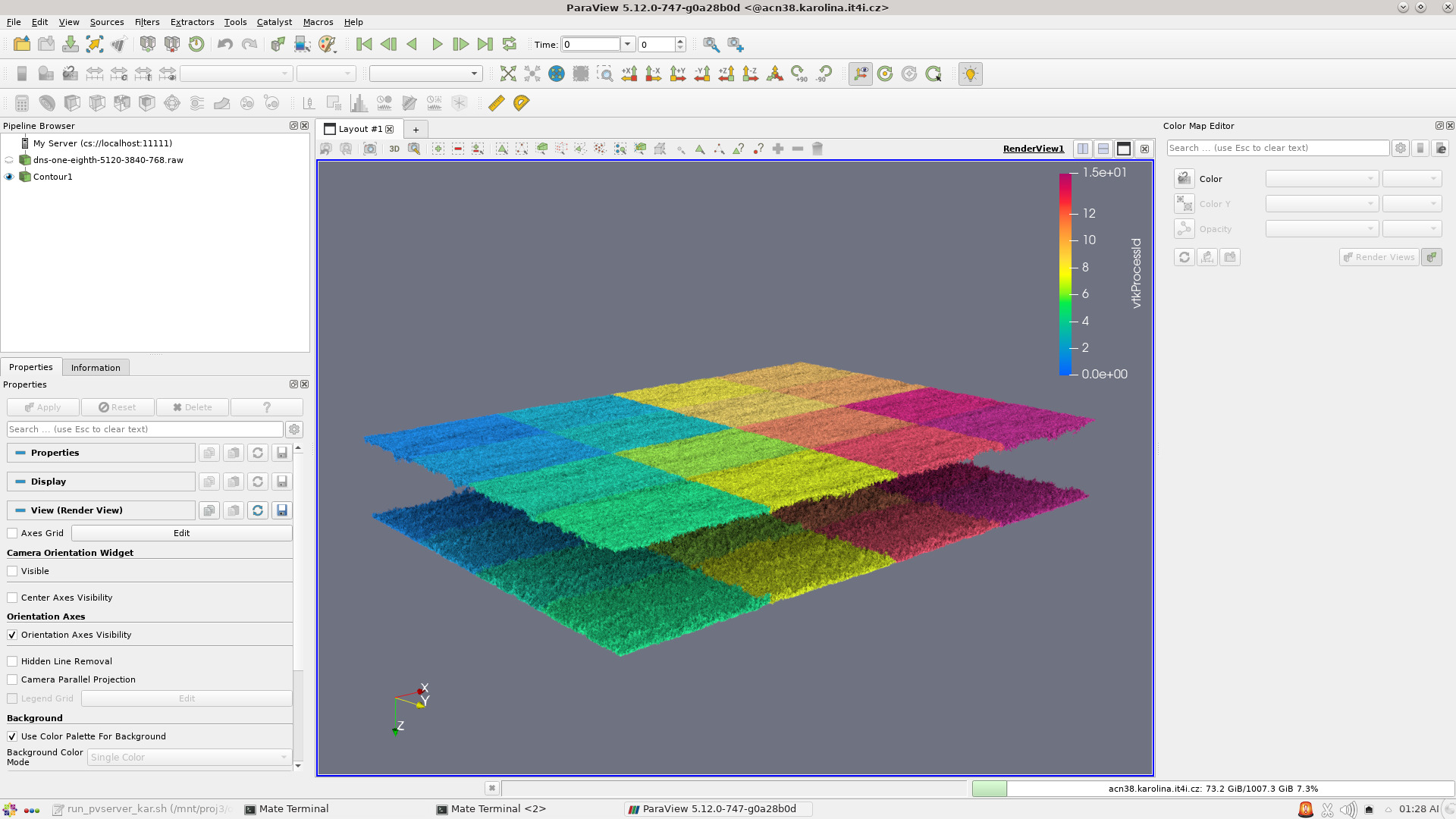}
  \includegraphics[width=.32\columnwidth]{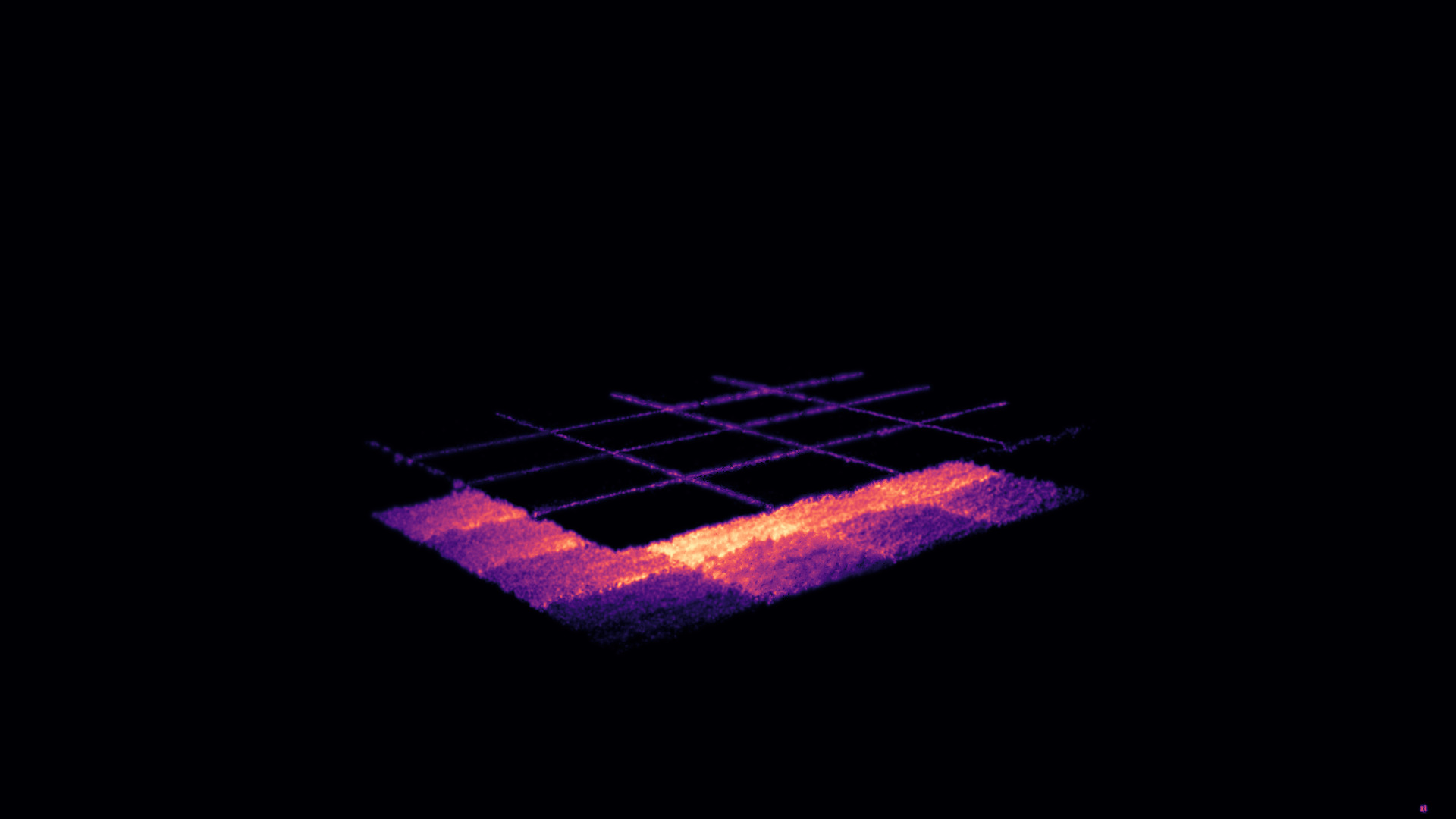}
\end{center}
\vspace{-2em}
\caption{\label{fig:composite-tests-diffuse} Capabilities and
  limitations of compositing: A test case of $4^3$ boxes
  pseudo-randomly interleaved across 4 ranks. Left, rendered with the
  pass-through compositing (PTC) device, compositing images from
  another ANARI device that does not know that other ranks (or their
  data) even exist. Our ANARI PTC device can properly z-composite
  these images, but cannot create (global) shadows. Middle, the same
  data-parallel application using a natively data-parallel ANARI
  device (Barney, in this case), with proper global shading effects.
  Right: difference images generated with \FLIP. Bottom row: similar
  experiment, conducted with the \code{DNS} data set in ParaView.
}
\end{figure}

\begin{figure}[ht]
  \setlength{\tabcolsep}{0pt}
  \begin{tabular}{cc}
    \hline
    \includegraphics[width=.495\columnwidth]{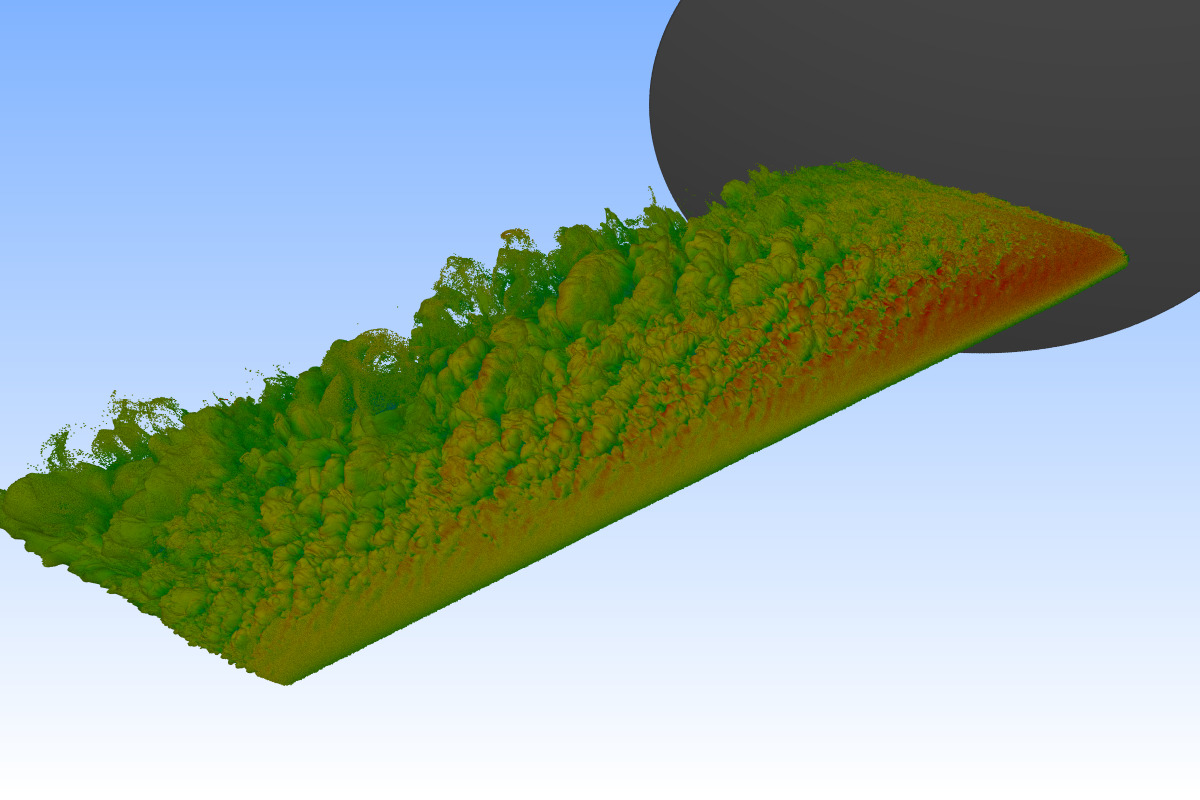}
    &
    \includegraphics[width=.495\columnwidth]{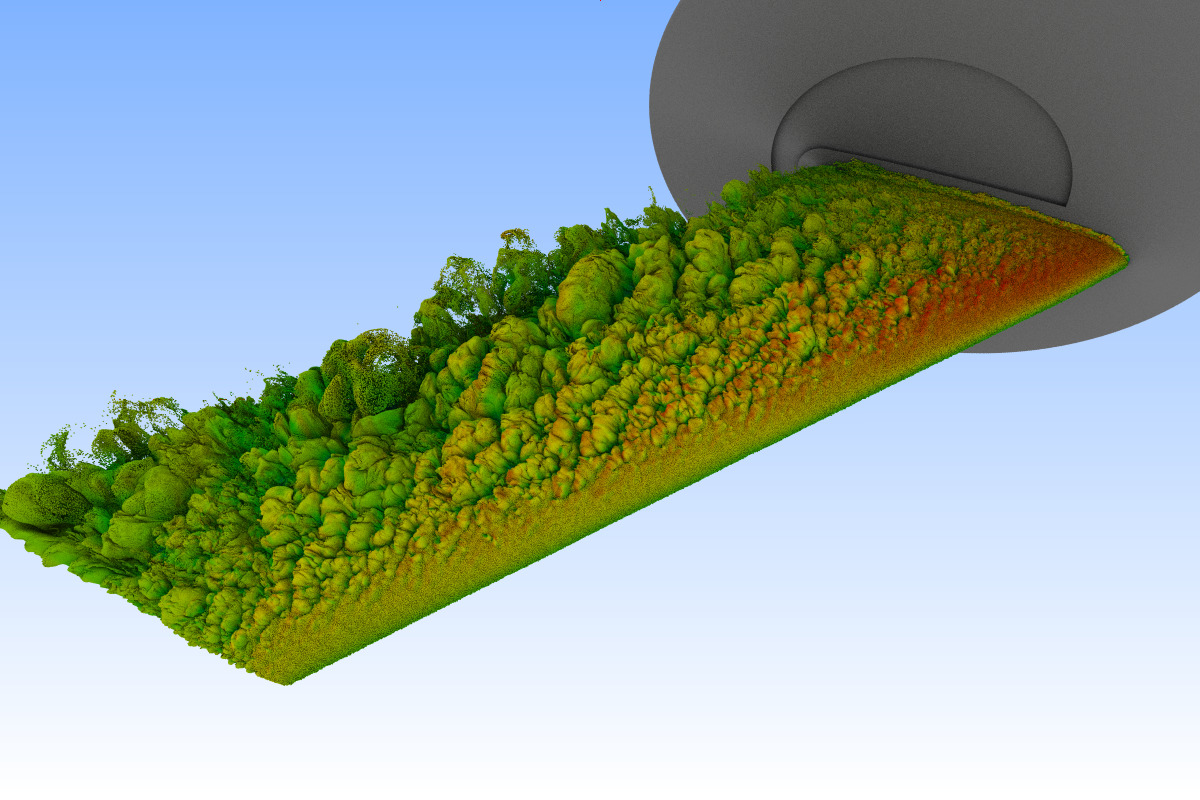}
    \\     
    \multicolumn{2}{c}{airflow around wing, ca 1 billion spheres}
    \\
    (local shading only)
    &
    (with path tracing)
    \\
    \hline
    \includegraphics[width=.495\columnwidth]{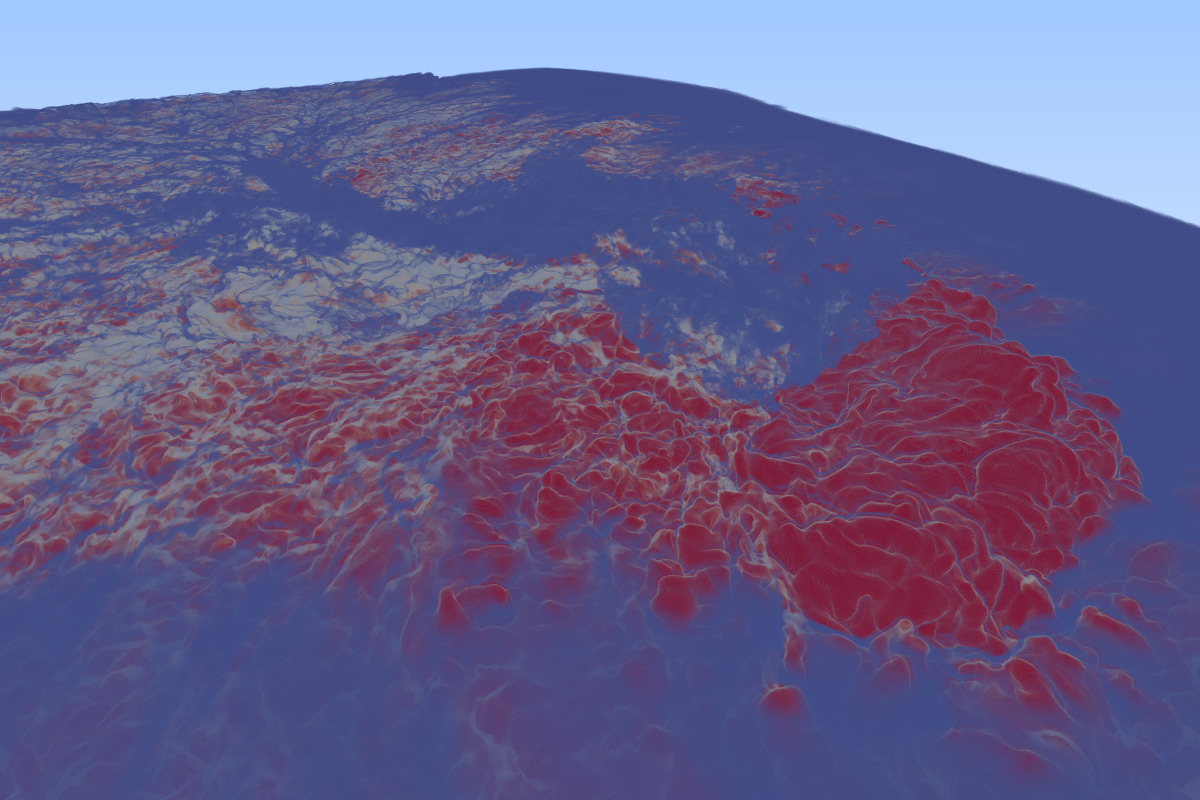}
    &
    \includegraphics[width=.495\columnwidth]{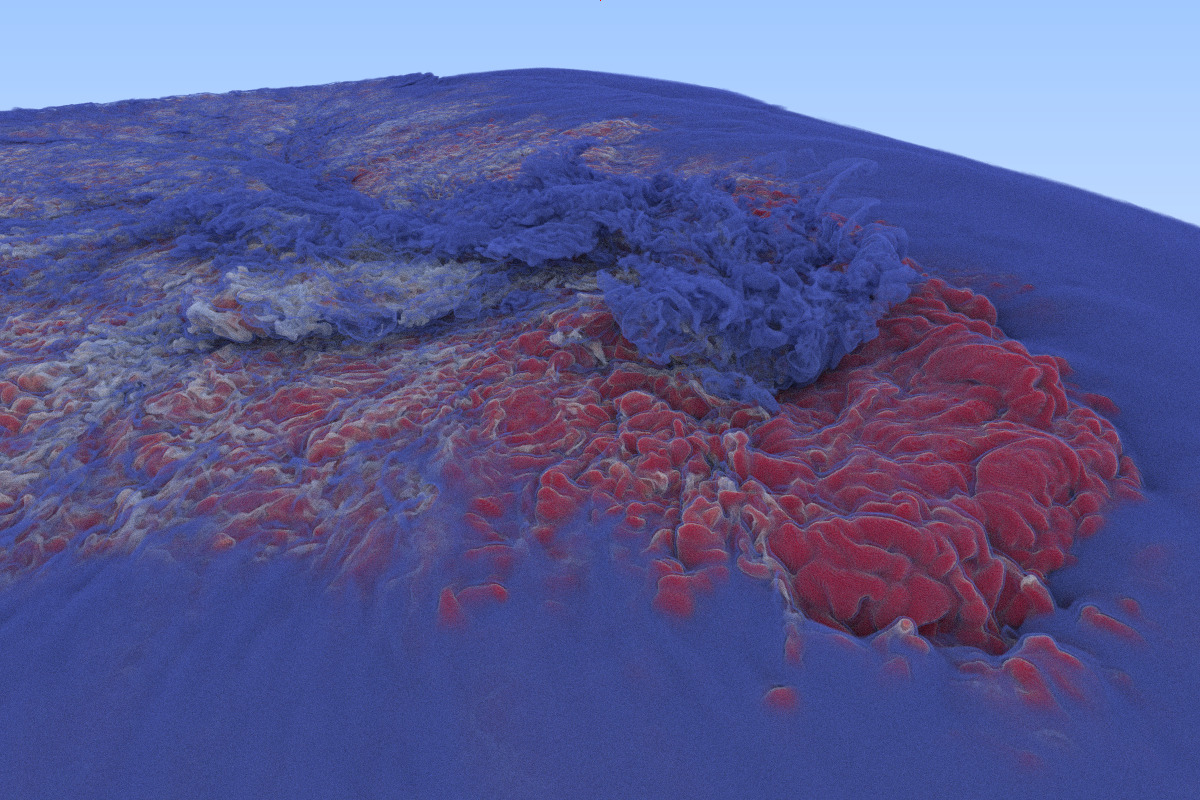}
    \\
    \multicolumn{2}{c}{thunderstorm data set (volume rendered)}
    \\
    (local-only shading model)
    &
    (with volumetric shadows)
    \\
    \hline
  \end{tabular}\\[-1em]
  \caption{Examples of why path traced rendering is important in the
    first place, even for data-parallel sci-vis. Top: a roughly 1
    billion spheres data set from NASA AMES (showing airflow around a
    wing). Bottom: the $2,501\times 3,001\times 305\times$\code{float}
    \emph{thunderstorm} data set. Left: with local shading only;
    right: with path tracing turned on. All images rendered with
    Barney, HayStack, and in data-parallel.
    \label{fig:before-after}
    \vspace*{-1em}
  }
\end{figure}

\added{Our motivation for a DP-ANARI paradigm is to enable global effects
usually realized with ray tracing in data-parallel sci-vis renderers.
Ordinary data parallel renderers used in sci-vis cannot produce these
effects without artifacts, as can be seen in \cref{fig:composite-tests-diffuse}.}
To illustrate the effect of missing global effects for some more
realistic data, in \cref{fig:before-after} we show two large
data sets (one volumetric, one surface based) rendered with Barney,
and in exactly the same data-parallel configuration---but once with
only local shading, and once with path tracing turned on. Clearly the
images with global effects are not only ``nicer'', but also convey
more information (which is what visualization is about)---but with classical
ANARI (i.e., without exposing the concept of a data parallel world) this
could not be reproduced. An application could of course still decide
to not use ANARI for such use cases at all, and instead integrate into
libraries like Barney in the first place---but this would lose all the
benefits of why applications are integrating the first place.

\begin{figure}[ht]
  \resizebox{1.00\columnwidth}{!}{
    \includegraphics[height=2.5cm]{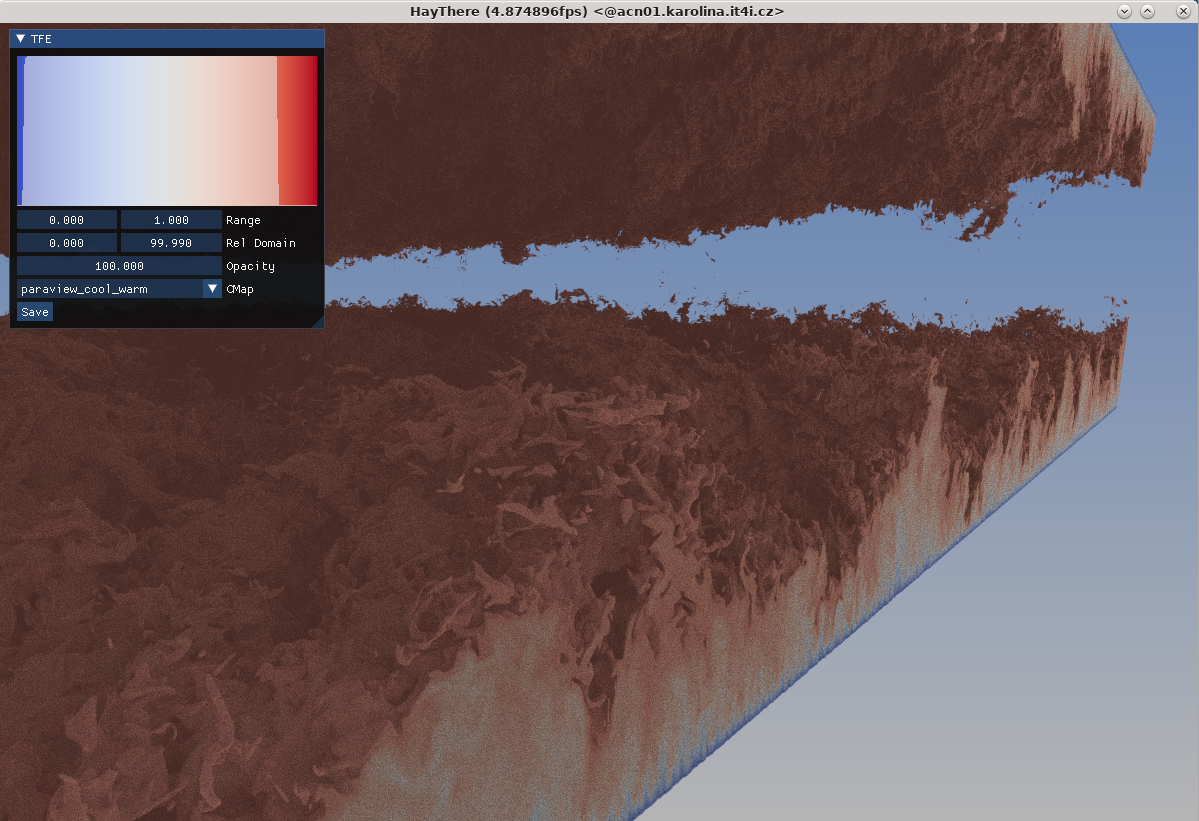}
    \includegraphics[height=2.5cm]{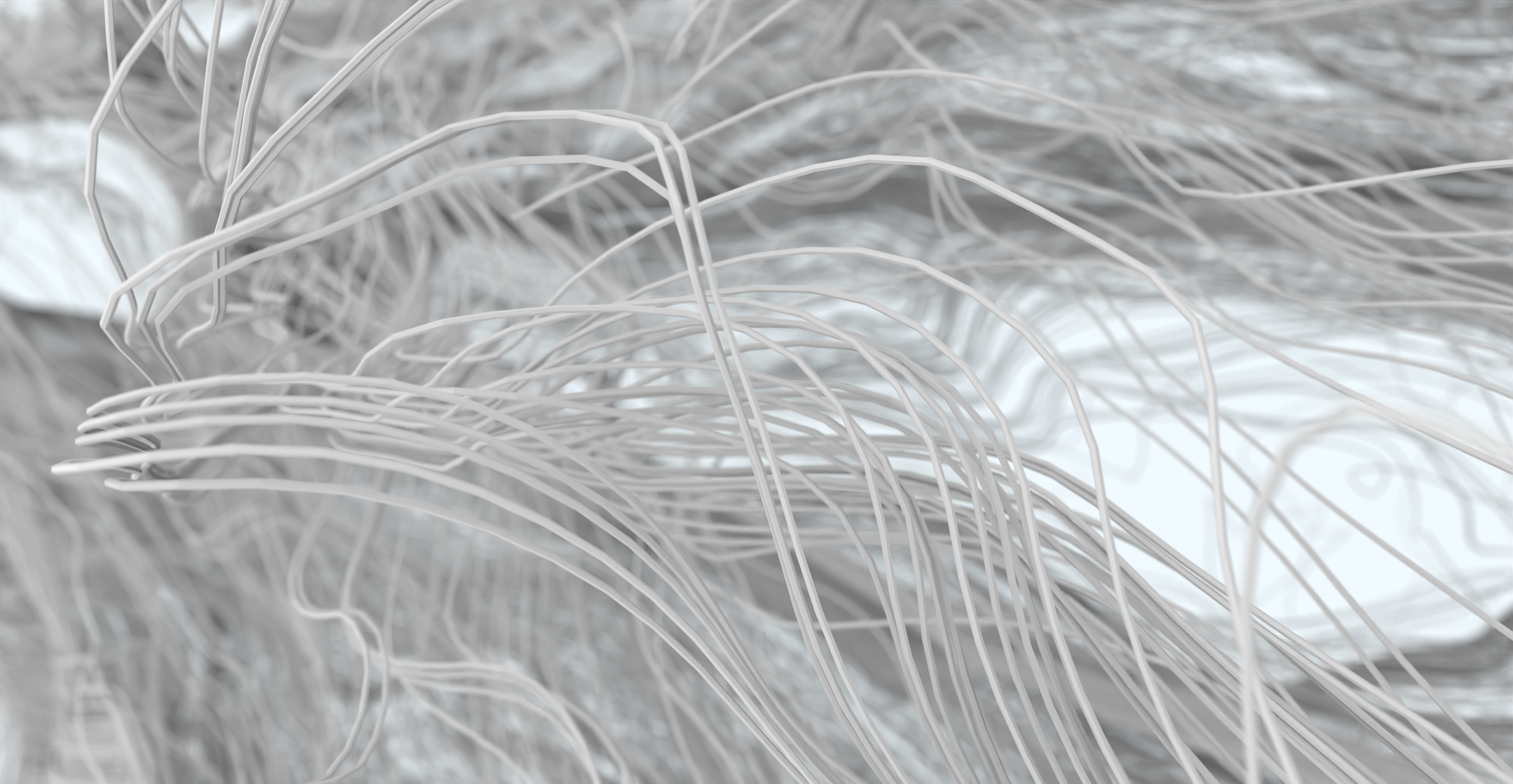}
  }\\
  \resizebox{1.00\columnwidth}{!}{
    \includegraphics[height=2.5cm]{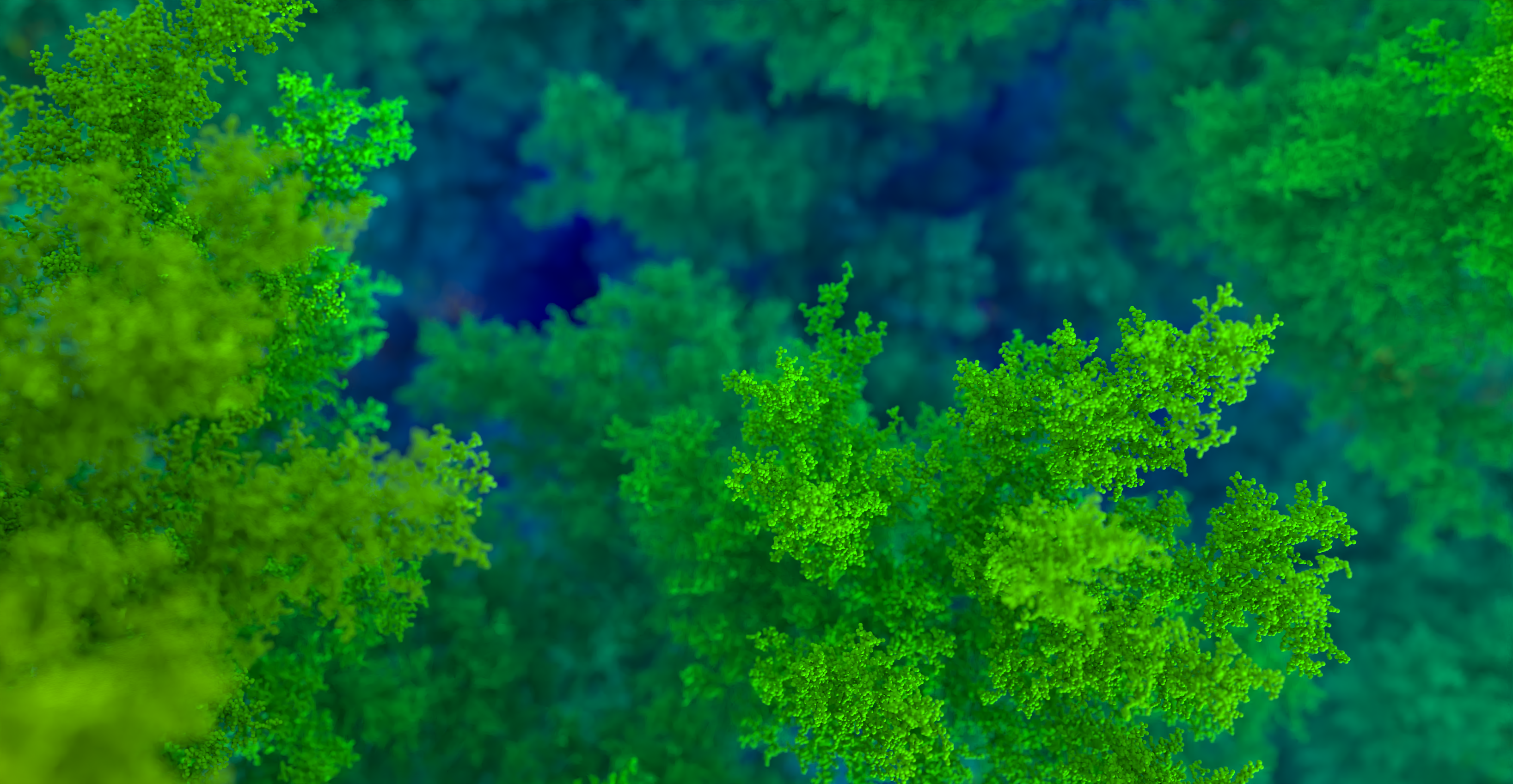}
    \includegraphics[height=2.5cm]{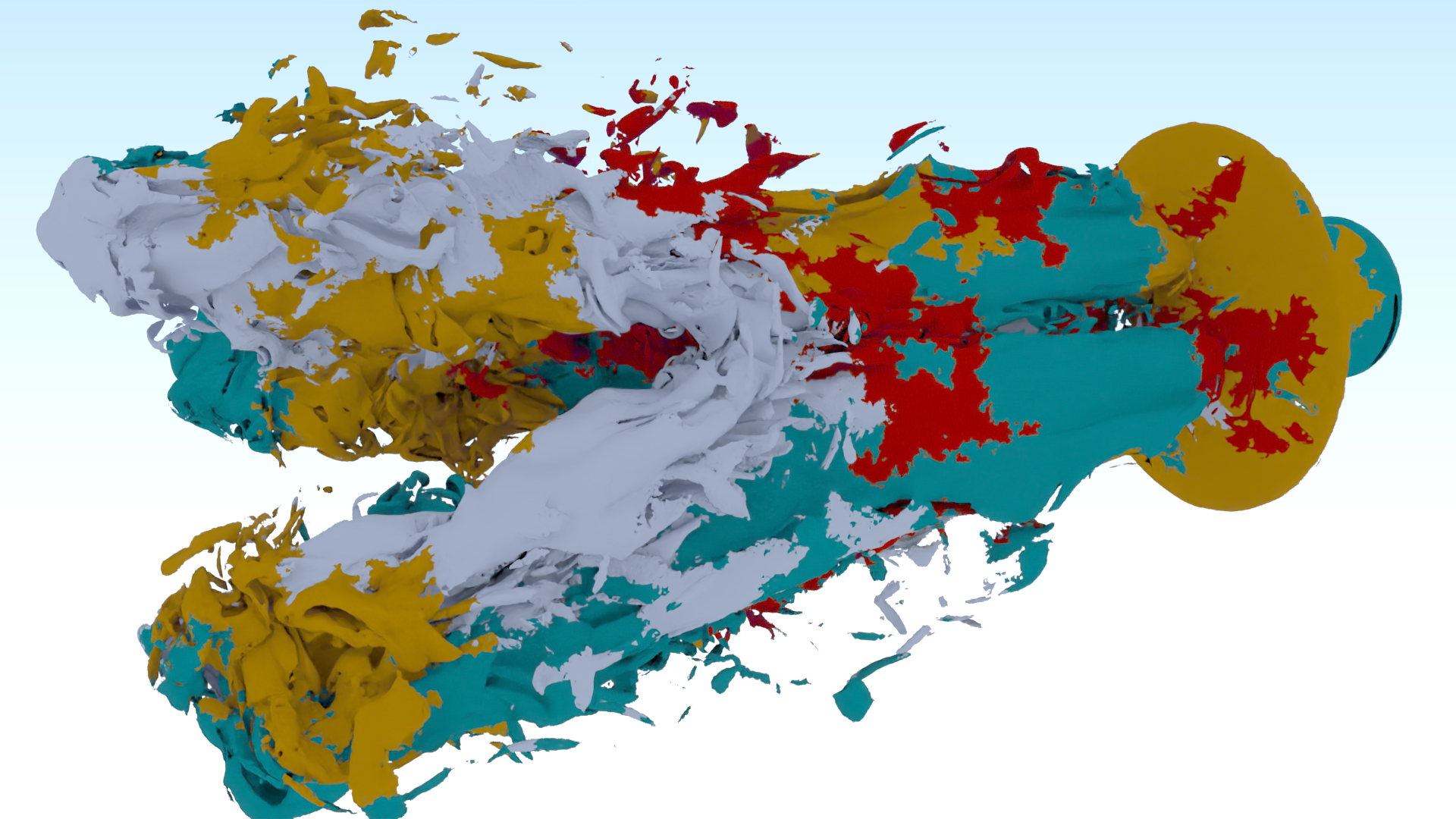}
  }\\[-4ex]
  \caption{Several examples of data-parallel path traced svi-vis
    data. Left to right, top to bottom: a) The $10,240\times
    7,680\times 1,536$ \code{DNS} data set, volume path traced on
    either RTX~8000 cards. b) streamlines from a jet flow simulation rendered
    with depth of field c) 250M particle DLA simulation also rendered with depth
    of field d)~A total of order 1 billion, extracted on-the-fly from the order
    1 billion element NASA Mars Lander dataset (see~\cite{big-lander}, and
    color-coded by which rank owned the underlying mesh part.
    \label{fig:more-examples}
    \label{fig:wow}
    \vspace*{-1em}
  }
\end{figure}

Whereas \cref{fig:before-after} explicitly illustrated
data-parallel rendering with and without advanced ray tracing effects,
in \cref{fig:teaser} and~\cref{fig:more-examples} we provide several more examples of
what an application could expect when adopting a data-parallel path
tracing paradigm.

\section{Discussion}

\removed{In this paper, we have proposed a paradigm for how the ANARI API could
be used for data-parallel rendering. We have shown that this can be
done without any new API calls, merely through formalizing a specific
workflow of how different ranks would \emph{jointly} issue their
respective API calls. Doing so allows an ANARI device to argue about
what these calls mean with respect to a \emph{global} ANARI world
that extends beyond just the given rank---which then allows this
device to do whatever it wants to do for rendering.}

\added{In this paper we have proposed a paradigm for using ANARI
for data-parallel rendering.}
We have not proposed any new method, nor a new
API, nor even a specific system. Instead, this paper should
be seen as an attempt to rally both developers and users of ANARI to
\emph{agree} on a specific way of using ANARI. Doing this is what we
believe to be the key to breaking the chicken-and-egg problem in which
--at least for data parallel rendering---vis applications are stuck
with compositing, while ANARI device developers cannot develop
data-parallel renderers because existing ANARI is purely single-rank.

We have formalized a workflow that is simple and flexible, yet
sufficiently expressive to work for both compositing and true
data-parallel path tracing. We have shown that it is quite easy to
implement this paradigm (for both of these categories), and have used
a variety of different (prototypical) integrations to show that this
is also easy to integrate. We have also shown that there is a very easy
on-ramp for applications to ease into this paradigm, by simply using
our (or any similar) compositing device---if this is used with
whatever ANARI device the app is currently using, the app would get
exactly the same outcome as before, while being able to also run any
true data-parallel path tracer when desired.

\subsection{Limitations}

The most obvious limitation of our approach is that it is not as
easily tangible as any new API extension would be. At its heart our
paradigm only specifies a \emph{convention}, and even that would
eventually need some sort of formalization in the ANARI spec. This is,
however, not all that different from how ANARI works in general: for
example, ANARI does the API call for creating a material with a given
name, but it does \emph{not} make any guarantees how (or even whether)
a given device will implement a given material. In practice, this
still works, due to what we call a \emph{normative pressure}: once
enough applications start to expect a given material to operate in a
given way, device developers come under significant pressure to
implement it the way that those applications expect.
We fully expect adoption of our paradigm to work exactly the same way.

Another limitation is that our paradigm specifies how a given scene is
to be created, but does not make any guarantees about how a given
device will then render it. This is yet another example of ANARI being
intentionally vague, and relying on said normative pressure; however,
we fully expect compositing-based devices to remain in use for a
considerable while, and applications will have to decide how to deal
with that.

\subsection{Scene/Data Partitioning}

One key issue for data parallel rendering---which we have completely
skirted so far---is how the scene is partitioned across the different
ranks. This is important because many data-parallel renderers will
only work for certain types of data partitioning. For example, IceT's
alpha blending mode~\cite{brix} requires a spatial partitioning of the
scene as well as an a-priori known compositing order; and similar
limitations would apply to other renderers. In Sci-vis, this problem
gets even more interesting because it is not the renderer that does
the scene partitioning, but the application---such as \code{pvserver}
for \code{ParaView}, or \code{libsim} for \code{VisIt}.

Clearly, if there is such a strong dependence on how the scene is
partitioned, any API or paradigm for data parallel rendering
\emph{must} have a means of communicating what the back-end can
consume, and/or what the front-end has generated. One way of solving
this would be to specify a certain partitioning requirement in the
API, but this would unduly restrict what kind of renderers could or
could not be implemented.

Instead, we suggest to handle that by having the apps pass such meta
information by setting a set of parameters on the underlying
device. For example, apps that do want to use devices that use IceT
could set some \code{int compositingOrder} and \code{box3 boundingBox}
parameter on the device. Of course, this only works if the app can
actually provide such data, but if it couldn't it wouldn't be able to
use IceT, anyway. Here, however, we would expect the
aforementioned normative pressure to eventually assert itself, too: if
some devices have more constraints than others then clearly these will
see some pressure to relieve these constraints.

\subsection{Remaining Issues}

The key remaining issue is to get the developers of actual tools like
ParaView and VisIt to adopt this paradigm. Though we believe this
paper to have made a strong argument that they should, this will
likely not happen over night.

Ultimately this will also require more work on the device
side. OSPRay, for example, already more or less follows the same
paradigm, and already has a (single-rank) ANARI interface---but would
yet have to merge these two. For our own devices, much is left to be
done, too: for Barney, there are still rather large gaps between what
Barney supports and what ANARI would expect. Adding these missing
features---and changing existing ones to be more ANARI-like---will
require significant effort. However, applications will not adopt it
until it supports enough of the ANARI features that said application
requires. This is another example of the aforementioned normative
pressure, but it will still not happen over night.

\section{Conclusion}

In this paper, we have proposed a paradigm---or convention---for how
data-parallel vis apps and data-parallel renderers can use the ANARI
API to jointly argue about a global scene. This by itself clearly does
not completely solve the problem of data-parallel rendering in either
sci-vis (nor even that of data-parallel rendering in ANARI). However,
we believe this paper to have made three major contributions towards
that goal: First, to have proposed what is essentially a road-map
towards true data-parallel path tracing in sci-vis rendering, which
both app and device developers can follow. Second, a set of arguments
\emph{why} app developers should join in this effort, and that there
is no longer a reason not to. And third, a set of devices, prototypes,
and proof-of-concepts that others can build on (all of which we have
made publicly available), and which we believe will be a foundation
for reaching a virtuous cycle where app developers and device
developers can now jointly work towards a common goal.


\bibliographystyle{abbrv-doi}

\bibliography{main}


\end{document}